\documentclass[journal]{IEEEtran}
\usepackage{cite}
\usepackage[acronym]{glossaries}
\usepackage{amsmath,amssymb,amsfonts}
\usepackage{listings}

\usepackage{graphicx}
\usepackage{booktabs}
\usepackage[referable]{threeparttablex}
\usepackage{multirow}

\usepackage[free-standing-units,per-mode=repeated-symbol,binary-units=true,detect-weight=true,detect-family=true]{siunitx}[=v2]

\usepackage[hyphens]{url}
\usepackage{cleveref}
\usepackage{xurl}

\usepackage[
monochrome
]{xcolor}

\usepackage{svg}
\usepackage{tikz}

\DeclareSIUnit\squaredkm{\si{km.s^{2}}}
\DeclareSIUnit\devspersquaredkm{dev/\si{km^{2}}}
\DeclareSIUnit\bps{bps}
\DeclareSIUnit\terabps{T\bps}
\DeclareSIUnit\gigabps{G\bps}
\DeclareSIUnit\megabps{M\bps}

\newacronym{lte}{LTE}{long-term evolution}
\newacronym{4g}{4G}{4th generation}
\newacronym{5g}{5G}{5th generation}
\newacronym{6g}{6G}{6th generation}
\newacronym{nr}{NR}{New Radio}
\newacronym{3gpp}{3GPP}{3rd Generation Partnership Project}
\newacronym{roi}{ROI}{return on investments}
\newacronym{ran}{RAN}{radio access network}
\newacronym{ru}{RU}{remote unit}
\newacronym{ue}{UE}{user equipment}
\newacronym{gnb}{gNB}{gNodeB}
\newacronym{pusch}{PUSCH}{Physical Uplink Shared Channel}
\newacronym{phy}{PHY}{physical layer}
\newacronym{tti}{TTI}{Transition Time Interval}
\newacronym{bs}{BS}{base station}
\newacronym{asic}{ASIC}{application-specific integrated circuit}
\newacronym{fpga}{FPGA}{field programmable gate array}
\newacronym{asip}{ASIP}{application-specific instruction processor}
\newacronym{gpp}{GPP}{general purpose processor}
\newacronym{vliw}{VLIW}{very-large instruction word}
\newacronym{cgra}{CGRA}{coarse-grain reconfigurable architecture}
\newacronym{cpu}{CPU}{central processing unit}
\newacronym{gpu}{GP-GPU}{general-purpose graphic processing unit}
\newacronym{cuda}{CUDA}{Compute Unified Device Architecture}
\newacronym{dsp}{DSP}{digital signal processing}
\newacronym{spmd}{SPMD}{single-program multiple-data}
\newacronym{isa}{ISA}{instruction set architecture}
\newacronym{rf}{RF}{register file}
\newacronym{ipu}{IPU}{integer processing unit}
\newacronym{fp}{FP}{floating-point}
\newacronym{fpu}{FPU}{floating point unit}
\newacronym{fpss}{FP-SS}{floating point sub-system}
\newacronym{simd}{SIMD}{single-instruction multiple-data}
\newacronym{lsu}{LSU}{load\&store unit}
\newacronym{numa}{NUMA}{non-uniform memory access}
\newacronym{tcdm}{TCDM}{Tightly-Coupled Data Memory}
\newacronym{dma}{DMA}{direct memory access}
\newacronym{icache}{I\$}{instruction cache}
\newacronym{xbar}{X-BAR}{crossbar}
\newacronym{raw}{RAW}{read after write}
\newacronym{war}{WAR}{write after read}
\newacronym{wfi}{WFI}{wait for interrupt}
\newacronym{ipc}{IPC}{instruction per cycle}
\newacronym{ber}{BER}{bit error rate}
\newacronym{snr}{SNR}{signal to noise ratio}
\newacronym{mac}{MAC}{multiply\&accumulate}
\newacronym{mimo}{MIMO}{Multiple-Input Multiple-Output}
\newacronym{mmse}{MMSE}{minimum mean squared error}
\newacronym{lse}{LSE}{least square estimation}
\newacronym{ofdm}{OFDM}{Orthogonal Frequency Division Multiplexing}
\newacronym{fft}{FFT}{fast Fourier transform}
\newacronym{che}{CHE}{channel estimation}
\newacronym{bf}{BF}{beamforming}
\newacronym{dmrs}{DMRS}{demodulation\&reference symbol}
\newacronym{mvm}{MVM}{matrix-vector multiplication}
\newacronym{matmul}{MMM}{matrix-matrix multiplication}
\newacronym{wdotp}{w-Dotp}{widening dot-product}
\newacronym{pnr}{PnR}{place\&route}
\newacronym{rtl}{RTL}{register transfer level}
\newacronym{ppa}{PPA}{power-performance-area}
\newacronym{soa}{SoTA}{state-of-the-art}
\newacronym{iq}{IQ}{in-phase and quadrature}
\newacronym{l1}{L1}{level-one}
\newacronym{l2}{L2}{level-two}

\newcommand\copyrighttext{\footnotesize \textcopyright This work was submitted to the IEEE for publication. Copyright may be transferred without notice, after which this version may no longer be accessible.}
\newcommand\copyrightnotice{%
    \begin{tikzpicture}[remember picture,overlay]
        \node[anchor=south,yshift=10pt] at (current page.south) {\fbox{\parbox{\dimexpr\textwidth-\fboxsep-\fboxrule\relax}{\copyrighttext}}};
    \end{tikzpicture}%
}

\begin{document}
%
\title{A \textcolor{red}{66Gbps/5.5W} RISC-V Many-Core Cluster for 5G+ Software-Defined Radio Uplinks}

%
%
%

\author{Marco~Bertuletti,~\IEEEmembership{Student Member,~IEEE,}
        Yichao Zhang,~\IEEEmembership{Student Member,~IEEE,}
        Alessandro~Vanelli-Coralli,~\IEEEmembership{Senior Member,~IEEE,}
        Luca~Benini,~\IEEEmembership{Fellow,~IEEE,}
\thanks{Marco Bertuletti and Yichao Zhang are with the Integrated Systems Laboratory (IIS), Eidegenossische Technische Hochschule (ETH), Zurich, Switzerland e-mail: mbertuletti@iis.ee.ethz.ch, yiczhang@iis.ee.ethz.ch.}
\thanks{Alessandro Vanelli-Coralli and Luca Benini are with the University of Bologna, Bologna, Italy, and with ETH e-mail: avanelli@iis.ee.ethz.ch, lbenini@iis.ee.ethz.ch.}
\thanks{}}

\markboth{}%
{Shell \MakeLowercase{\textit{et al.}}: Bare Demo of IEEEtran.cls for IEEE Journals}

\maketitle

\copyrightnotice
\begin{abstract}
Following the scale-up of New Radio complexity in 5G and beyond, the physical layer's computing load on base stations is increasing under a strictly constrained latency and power budget: base stations must process \textgreater~\si{20~Gbps} uplink wireless data rate on the fly, in \textless~\si{10~W}. At the same time, the programmability and reconfigurability of base station components is a key requirement: it reduces the time and cost of new networks' deployment, it lowers the acceptance threshold for industry players to enter the market, it ensures return on investments in a fast-paced evolution of standards. In this paper, we present the design of a many-core cluster for 5G and beyond base station processing. Our design features \num{1024}, streamlined RISC-V cores with domain-specific floating-point extensions, and \si{4~MiB} shared memory. It provides the necessary computational capabilities for software-defined processing of the lower physical layer of 5G physical uplink shared channel (PUSCH), satisfying high-end throughput requirements (\si{66~Gbps} for a \gls{tti}, \si{\num{9.4}\mbox{-}302~Gbps} depending on the processing stage). 
The throughput metrics for the implemented functions are 10 times higher than in state-of-the-art application-specific instruction processors. The energy efficiency on key New Radio kernels (\si{2\mbox{-}41~Gbps/W}), measured at \si{800~MHz}, \num{25}$^\circ C$, \si{\num{0.8}~V}, on a placed and routed instance in 12nm CMOS technology, is competitive with state-of-the-art architectures. The PUSCH processing runs end-to-end on a single cluster in \si{\num{1.7}~ms}, at \si{<6~W} average power consumption, achieving \si{12~Gbps/W}.
\end{abstract}


\glsreset{tti}

\IEEEpeerreviewmaketitle

\section{Introduction}
\IEEEPARstart{T}~he \gls{5g} wireless system standard has been a key enabler for high data rate, low latency, and ultra-reliable mobile services. It was introduced in 2017 by the \gls{3gpp} Release-15~\cite{3GPP_TS_R15}, and it enabled numerous applications: augmented reality, internet of things, device-to-device and vehicles to everything communications, remote healthcare, machine-to-machine and drone communications~\cite{Agiwal_COMST_2018, Ullah_5Gusecases_IEEE_2019}. Application studies expect that transitioning to \gls{6g} will bring further advancements, including telepresence, autonomous driving vehicles, indoor and outdoor wireless positioning\&sensing~\cite{Qadir_6Gusecases_2023, Shen_6Gresearch_ACM_2022, Rappaport_THz_IEEE_2019}. The total global mobile data traffic, reached \si{145~EB/month} at the end of 2023 and is projected to grow by \si{\num{3.5}\times}, touching the threshold of \si{466~EB/month} in 2029 \cite{Ericsson_MobilityReport_2024}. 

As a consequence of this staggering data volume, computing challenges are rapidly rising at the edge nodes of the telecommunication infrastructure. The key requirements for \gls{5g} and \gls{6g} networks mandate that the peak uplink data rate will grow from \si{\num{20}~Gbps} to \si{\num{1}~Tbps}, and the round trip latency will decrease, from \si{\num{1}~ms} to \si{\num{0.1}~ms}. There will be \si{10\times} more devices per unit area (from \si{10^{6}}{\devspersquaredkm} to \si{10^{7}}{\devspersquaredkm}), and the spectrum usage will extend from below \si{\num{6}~GHz} up to the sub-\si{THz} frequency bands~\cite{ITU_2017, Shen_6Gresearch_ACM_2022, Rappaport_THz_IEEE_2019}. Therefore, extremely high-performance \gls{dsp} hardware must be developed to sustain huge data rates at an acceptable throughput. Optimizations will be focused on the \glspl{ru} of \glspl{ran} because the up-link and down-link protocols implemented in the \glspl{ru} directly affect the \glspl{ue} experienced latency and require the fast consumption of data on the flight~\cite{Larsen_COMST_2019}. 

Nowadays, digital processing in \gls{5g} \glspl{bs}, i.e. the digital part of \gls{gnb} processing, is performed on heterogeneous platforms. General-purpose programmable cores orchestrate the user data streams, while different \gls{asic} accelerators process the signal~\cite{Ceva_PentaGRAN}. Looking forward, in the competitive landscape of a fast-paced evolution telecommunication standard, resorting to fully programmable software-defined \glspl{bs} is a highly desirable paradigm shift. First, the software-defined paradigm~\cite{Mitola_SDR_IEEE_1995, Machado_SDR_IEEE_2015}, allows reducing the time to market in the deployment of new telecommunications standards, by softwarizing the blocks of the signal chain that can most rapidly evolve and by opening the way for differentiation opportunities and value-added functions. Many industry players adapted to this trend, offering programmable hardware \gls{bs} components and software libraries tailored for them \cite{Marvell_SDK, Quallcom_SDK}. Second, a software-defined approach helps to lower startup costs, increasing productivity and market growth. Third, it reduces the need to replace components in an installed \gls{bs}, while making sure it can keep up with the evolving standard. The latter is a key economic motivation for companies to move towards softwarization of their processing: most of the vendors estimate the \gls{roi} over wireless networks deployment to 5-10 years\cite{Huawei_ROI}, but telecommunication protocols evolved faster. Between 2015 and 2025 they moved from \gls{lte}, to \gls{4g}, then to \gls{5g}, and finally to \gls{6g}.

Energy efficiency is the key metric to trade off with programmability in software-defined \gls{gnb} processing. Upper bounds on power consumption emerged upon pushing the \gls{phy} processing to the \glspl{ru} edge network nodes, to improve the quality of service: \textcolor{red}{nowadays, the \glspl{ru} contribute to \si{40\%} of total \gls{ran} power consumption~\cite{Wesemann_IEEE_2023}. According to a survey on \gls{4g} networks~\cite{Blume_IEEE_2010} up to \si{90\%} of the  \si{1\mbox{-}2~kW} consumed by a \gls{5g} \gls{bs} with 64 transceivers~\cite{Huawei_Power} is dissipated in essential operational costs (power amplifiers, power supply, and air conditioning). With only \si{100~W} left for analog and digital signal processing, we can assume \si{10~W} power consumption per \gls{bs} component. Therefore, beyond \gls{5g} processors require \si{> 2~Gbps/W} energy efficiency.}

In this paper, we move from the general architectural template of TeraPool~\cite{Zhang_TeraPool_2024}, a fully programmable \num{1024}-core cluster, and we enhance and develop it for \gls{gnb} signal processing, in particular for the lower-\gls{phy} of the 5G-\gls{pusch} channel. Our many-core design is fundamentally different from legacy \gls{gnb} processing solutions based on inflexible \gls{dsp} accelerators \cite{Peng_MMSE_ASIC_TCAS_2018, Tang_MMSE_ASIC_JSSCC_2021, Shahabuddin_MIMO_TVLSI_2021, Guo_FFT_ASIC_TVLSI_2023, Yang_FFT_ASIC_TCAS_2023, Zhang_MMSE_ASIC_ISSCC_2024}. Our design matches the requirements of software-defined wireless processing: each core of the compute fabric can be individually programmed. Our solution also addresses the challenges created by programmable clusters with few big cores, which would suffer from work imbalance and data movement overhead \cite{EdgeQ, Picocom_PC802, Marvell_Octeon10, Quallcom_X100}. We propose instead an architecture with hundreds of small, energy-efficient cores that access shared data memory in parallel. With this architecture, we enable true \gls{spmd} on a large workload generated by the lower-\gls{phy} \gls{pusch} processing. 

The micro-architecture of our cores is based on the RISC-V open \gls{isa}. We open-sourced our \gls{rtl} code, toolchains, and software\footnote{https://github.com/pulp-platform/mempool}, promoting hardware-software co-design, in line with the software-defined approach.

This work represents a significant extension of our previous exploration on the efficient parallelization of \gls{5g}-\gls{phy} micro-kernels on many-core clusters\cite{Bertuletti_PUSCH_2023}. The key contributions of this paper are summarized as follows:
\begin{enumerate}
    \item We design extensions for the RISC-V cores to support \gls{fp} computation at low hardware overhead (Zfinx~\cite{RISCV_zfinx}), augmented with domain-specific instructions to address the most common formats and precisions of telecommunication data (fixed-point or \gls{fp} complex \si{16~b} real, \si{16~b} imaginary).
    \item We implement software-defined end-to-end processing for the lower-\gls{phy} of \gls{5g} \gls{pusch} (i.e. demodulation, beamforming, channel estimation, and signal detection), one of the most demanding tasks of the \gls{phy} at the \gls{bs}. We schedule the processing steps to promote data reuse in the \gls{l1} memory, reducing the runtime of \si{12~\%} compared to loading all operands and storing all outputs in \gls{l2} memory for each step.
    \item We place and route our design in a \si{12~nm} FinFET technology, and measure its throughput and energy efficiency at \si{800~MHz}, \num{25}$^\circ C$, \si{\num{0.8}~V}, in a high-load use case. We compare our results with \gls{soa} solutions for \gls{gnb} signal processing.
\end{enumerate}
We estimate average power consumption of \si{<6~W} and a total latency of \si{\num{2.6}~ms} (fixed-point) and \si{\num{1.7}~ms} (\gls{fp} plus domain-specific extensions) for the execution of the addressed \gls{pusch} functions, at \si{12~Gbps/W} per symbol, which is highly competitive with all the \gls{soa} solutions for which we could gather performance, power and efficiency data.

The remainder of this paper is organized as follows: \Cref{sec:TeraPool_architecture} describes the architecture and programming model of the many-core cluster developed to sustain the high throughput requirements of beyond \gls{5g} workloads, focusing on the processing cores architecture and \gls{isa} and the hierarchical core-to-memory interconnect. \Cref{sec:PUSCH_implementation} describes: \textcolor{red}{III-A the uplink workload selected to benchmark the performance of the cluster; III-B the implementation of \gls{5g} parallel micro-kernels on the cluster, and III-C their performance; III-D the full \gls{pusch} pipeline execution}. \Cref{sec:results} presents the physical implementation and \gls{ppa} of our many-core cluster on software-defined lower-\gls{phy} workloads, introducing a comparison on energy efficiency with \gls{soa} \glspl{asip} and \gls{asic} in \Cref{sec:related_works}. \Cref{sec:conclusions} concludes the paper.

\section{TeraPool for software-defined physical layer: Architecture \& Programming Model}
\label{sec:TeraPool_architecture}

We address the low latency and high throughput of \gls{5g} \gls{nr}, exploiting the performance and scalability of TeraPool~\cite{Zhang_TeraPool_2024}. This large cluster features \num{1024} processing cores, each supporting the RV32IMAF \gls{isa}, and a shared multi-banked \gls{l1} scratchpad.
This section details the micro-architecture of the cores and the \gls{isa} extensions introduced for software-defined \gls{phy} workloads.
We describe TeraPool's shared memory interconnect, focusing on the advantages offered in simplifying the programming model and reducing data transfers. Finally, we give details of the \gls{spmd} programming model adopted for software-defined \gls{phy} kernels.

\subsection{Core-Complex: Physical Layer Specific Extensions }

\begin{figure}[ht]
  \centering
  \includegraphics[width=\columnwidth]{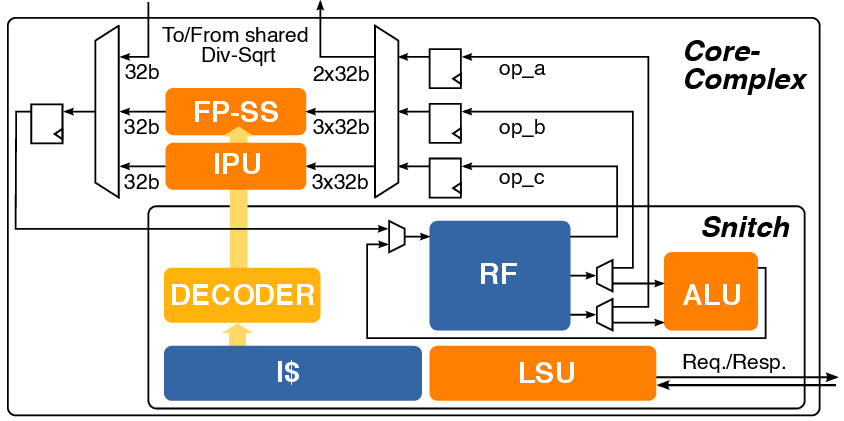}
  \caption{Snitch core with offload ports to the IPU, FP-SS, and shared division and square-root functional units. According to Zfinx specifications, the FP operands are in the integer RF.}
  \label{fig:terapool_corecomplex}
\end{figure}

The TeraPool core-complex, represented in~\Cref{fig:terapool_corecomplex}, is based on the \textit{Snitch} single-stage, single-issue lightweight core~\cite{Zaruba_snitch_2021}. The core decodes and executes instructions in the RV32I base \gls{isa} in one cycle, it offloads multi-cycle instructions to specialized functional units. Snitch book-keeps the dependencies of offloaded instructions in a scoreboard and proceeds in the program flow. As a result, their additional latency is hidden as long as they do not have data dependencies~\footnote{Snitch can keep fetching and executing new instructions if their input operands do not depend on operands of previously issued and not yet completed instructions.}. The Snitch core-complex in TeraPool includes an \gls{ipu}, implementing integer multiplication and division, and the \textit{Xpulpimg} extensions. The latter provide \gls{simd} shuffling, packing, and unpacking of half types and are useful to manipulate complex \si{16~b} data, typical in classical radio signal processing.

\begin{figure}[ht]
  \centering
  \includegraphics[width=\columnwidth]{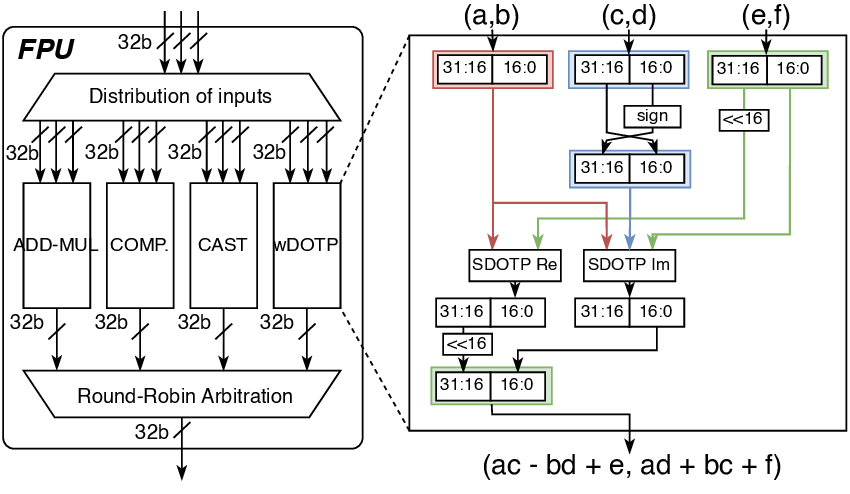}
  \caption{The modular FPU circuit has four functional slices. The wDotp unit is extended to perform one-cycle complex wDotp.}
  \label{fig:terapool_fpu}
\end{figure}

For \gls{pusch} processing and more in general for \gls{5g} (and beyond) software-defined \gls{phy}-processing, we also included \si{32~b} and \si{16~b} \gls{fp} support. In \gls{gnb} workloads, \gls{fp} numbers help code development: they reduce the full-scale range adjustments, required to combine processing steps and to fine-tune the application if the transmission context changes.
They also help to keep high numerical precision for the inversion of large matrices, which is a common operation for \gls{mimo} antenna systems~\cite{Ingemarsson_inversion_2015}.

\textcolor{red}{To maximize the compute units' area-efficiency, we support the \textit{Zfinx} and \textit{Zhinx} standard extensions~\cite{RISCV_zfinx}. They implement the same \gls{fp} 32b and 16b instructions as in the \textit{F} and \textit{Zfh} extensions, but use only the 32 registers available in the base core (also used for integer operations), eliminating the area overhead of a \gls{fp}-\gls{rf} and dedicated \gls{fp} load\&stores.} The \gls{fpss} shares the same port of the \gls{ipu} and includes a decoding stage and the configurable \gls{fpu} \textit{fpnew} module~\cite{Mach_fpnew_2021} in \Cref{fig:terapool_fpu}. This component groups and implements \gls{fp} extensions in four slices: addition\&multiplication, comparison, cast, and dot-product.

To further improve the cores' performance on \gls{5g} kernels we configure the \gls{fpu} for \gls{simd} \si{16~b} and \gls{wdotp}~\cite{Bertaccini_Minifloats_2024} support. We exploit the \gls{wdotp} \si{16~b} to \si{32~b} slice of the \gls{fpu} to implement \gls{fp} arithmetic on \si{32~b} complex data types (\si{16~b} real, \si{16~b} imaginary part). We instantiate one \gls{wdotp} slice for the real part and one for the imaginary. The \si{16~b} input and output operands are combined as in ~\Cref{fig:terapool_corecomplex}.

The receiving step of a \gls{5g} \gls{gnb} often requires matrix-inversions~\cite{Shahabuddin_MIMO_NorCAS_2020}. \gls{fp} division and square-root instructions are often used to implement these operators. Four cores share the same division and square-root unit with a round-robin policy to reduce the area overhead of these extensions, which are typically less frequently used than the basic \gls{fp} operators.

Each core complex has a private \gls{lsu}. Loads, stores, and atomic instructions support are discussed in the next subsection, alongside the cluster \gls{l1} interconnect.

\begin{figure*}[ht]
  \centering
  \includegraphics[width=\linewidth]{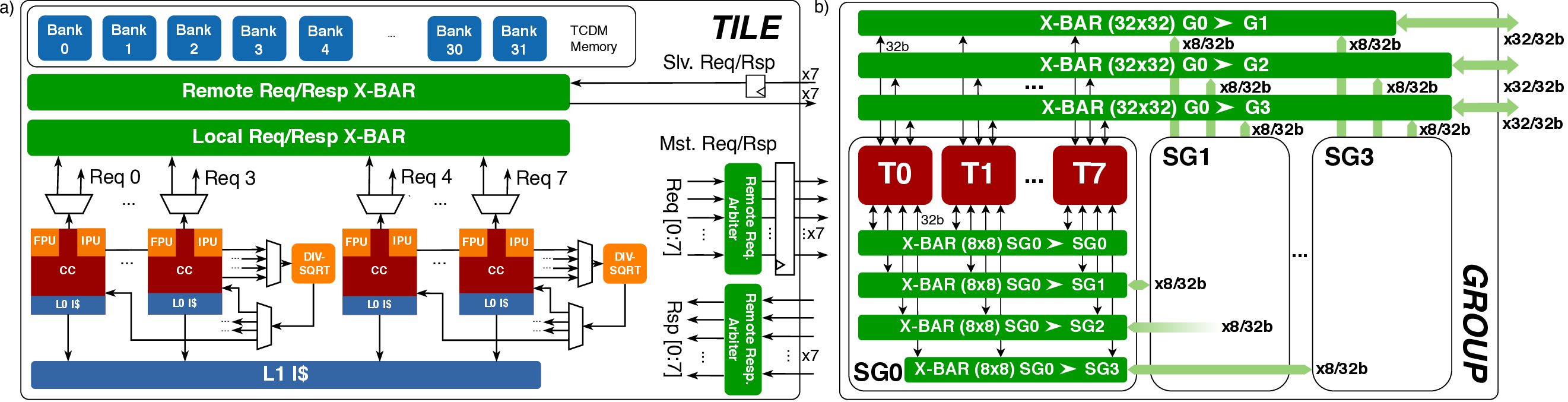}
  \caption{a) The TeraPool Tile, with 8 Snitch core-complexes and 2 shared division and square-root \gls{fp} units. The local Req/Resp X-BAR gives 1 cycle access to the shared TCDM, 7 remote Req/Resp master, and slave ports connect to other hierarchical levels. b) The connections between TeraPool Tiles (T0-7 for each SubGroup) and SubGroups (SG0-3) in a Group.}
  \label{fig:terapool_interconnect}
  \vspace{-1.5em}
\end{figure*}

\subsection{Cluster Interconnect}

In TeraPool, the \gls{l1} memory hierarchy is a multi-banked scratchpad, fully shared between \num{1024} processors. This architecture can be programmed with a streamlined \gls{spmd} approach, where all the cores have low latency access to data structures in \gls{l1} memory, that they process in parallel. However, routing congestion and the related effort in floorplanning, placement, and routing make an all-to-all flat \gls{xbar} interconnect between the cores and the scratchpad banks not physically feasible, even for smaller configurations than TeraPool's~\cite{Riedel_MemPool_2023}. To ensure physical feasibility, the cores are partitioned into \num{4} hierarchical levels. The cross-hierarchy memory access paths are shared by groups of cores with round-robin arbitration. Pipeline cuts are introduced on these paths to achieve near-\si{GHz} frequency targets, resulting in a \gls{numa} latency shared-\gls{l1} architecture~\cite{Zhang_TeraPool_2024}. There is a large configuration space in this \num{4}-level interconnect hierarchy. For software-defined processing of the lower-\gls{phy}, we select the most promising configuration for energy-efficiency~\cite{Zhang_TeraPool_2024}. In a low load traffic condition, any core in the cluster accesses any bank in less than \num{9}-cycles.

The \textit{Tile} in \Cref{fig:terapool_interconnect}a is the \textit{1st} implementation hierarchy. It contains \num{8} core complexes, \num{2} shared division and square-root \gls{fp} units, \si{4~KiB} of shared \gls{icache}, and \si{32~KiB} of scratchpad memory, divided into \num{32} banks, namely the \gls{tcdm}. The cores' \glspl{lsu} access the banks in \num{1}-cycle through a local \gls{xbar} built with a logarithmic number of stages. A reservation table per bank handles atomic load\&stores. Cores can hide TeraPool's \gls{numa} interconnect latency by issuing up to \num{8} outstanding transactions. Issued load\&stores are tracked by keeping their metadata in a transaction table. Unless a \gls{raw} dependency occurs, load\&stores are non-blocking for the next instruction execution.

\Cref{fig:terapool_interconnect}b represents the hierarchical connections of Tiles that we adopted for software-defined lower-\gls{phy} applications. The \textit{2nd} implementation hierarchy is a \textit{SubGroup}, with \num{8} Tiles. All cores in a Tile have a shared request-response port to an \num{8}$\times$\num{8} \gls{tcdm} \gls{xbar}, addressing the memory of other Tiles in the SubGroup. 
The \textit{3rd} implementation hierarchy is a \textit{Group}, with \num{4} SubGroups. In a Group, the Tiles of a SubGroup access the \gls{tcdm} of Tiles in other SubGroups via a shared request-response port, connected to one \num{8}$\times$\num{8} \gls{tcdm} \gls{xbar} per SubGroup.
The \textit{4th} level of the hierarchy is a \textit{Cluster}, with \num{4} Groups. The \num{32} Tiles in a Group access the \gls{tcdm} of Tiles in other Groups via \num{32}$\times$\num{32} \gls{tcdm} \glspl{xbar}, and one shared request-response port per Group in each Tile.

As described in \cite{Riedel_MemPool_2023}, an AXI interconnect is also instantiated in each hierarchical level. The AXI interconnect has three main functions. 
First, it connects the cores to the \gls{l2} memory, the cluster peripherals, and a cluster-shared \gls{dma} engine.
Second, it allows \gls{icache} refill. The cores and the \gls{icache} share the same AXI port at the Tile level.
Third, it carries streamlined \gls{dma} transfers between the \gls{l2} and the shared-\gls{l1} scratchpad banks. A \gls{dma} frontend is individually programmed by each core through reads and writes on the AXI interconnect. Data transfers are initiated by the frontend. They run over the AXI interconnect and they are redistributed to banks by a \gls{dma} backend in each SubGroup. The \gls{dma} backend and the Tiles' AXI requests and responses share the same \si{512~b} wide port at the SubGroup level. Therefore, the whole cluster has a \si{\num{1024}~B/cycle} AXI link to \gls{l2} memory, containing both instructions and data. 

\subsection{SPMD Programming Model for Software-Defined \gls{phy}}
The big workloads of \gls{5g} lower-\gls{phy} are parallelized over the many cores of the cluster, using a fork-join programming model. By default, all the cores execute the program in parallel. By runtime calls, each core reads its private ID from a status-register. The programmer can use this ID to index conditional branches and loops, distributing the workload between cores. At the end of the assigned parallel task, cores synchronize and enter a new parallel task.
Adopting the same \gls{spmd} paradigm used for small \gls{tcdm} clusters of \num{4} to \num{16} cores has the advantage of simplifying the programming model, but it could incur synchronization overheads. The cluster's C-runtime includes two main synchronization primitives~\cite{Bertuletti_Barriers_2023}. In the \textit{linear barrier}, cores atomically write a shared synchronization variable and go in a \gls{wfi} state; the last core fetching from the shared memory location resets the variable and wakes up all the others. The \textit{logarithmic barrier} implements a synchronization tree: first, the cores synchronize in groups by atomic writes to multiple locations, then the last core in each group continues to the next level of the tree, where the same pattern repeats. The core passing through all the synchronization levels wakes up the cluster. The wake-up is centrally handled, via a shared wake-up register and hardwired wake-up trigger connections to each core. The wake-up triggers to each core can be asserted with a Tile granularity, implementing \textit{partial synchronization} between subsets of cores in the cluster~\cite{Bertuletti_Barriers_2023}.

\textcolor{red}{We developed an open-source library of wireless \gls{dsp} kernels deeply optimized for our many-core architecture. The kernels are written in C and compiled with \textit{LLVM 12.0}. We modified the compiler backend to support the added \gls{phy}-processing \gls{isa} extensions.} For the design of kernels, we adopt a bare-metal approach. We parallelize loops assigning independent time iterations or input data portions to different cores, according to their unique ID. Kernels can be run standalone, using the compute fabric as a programmable accelerator, or stitched together, to set up an in-line processing pipeline. This is further clarified in \Cref{sec:PUSCH_op} and \Cref{sec:PUSCH_scheduling}.

\section{\gls{pusch} Computation}
\label{sec:PUSCH_implementation}

This section describes \gls{pusch}, one of the most computing-intensive channel of the \gls{5g} uplink. We detail the parallelization scheme adopted to implement the \gls{pusch} addressed functions and present an execution of the whole processing chain, up to the \gls{mmse} operation, that minimizes data-transfers between the shared-\gls{l1} and the upper layers of the memory hierarchy.

\subsection{\gls{pusch} Processing Steps}
\label{sec:PUSCH_processing}

\begin{figure}[ht]
  \centering
  \includegraphics[width=\columnwidth]{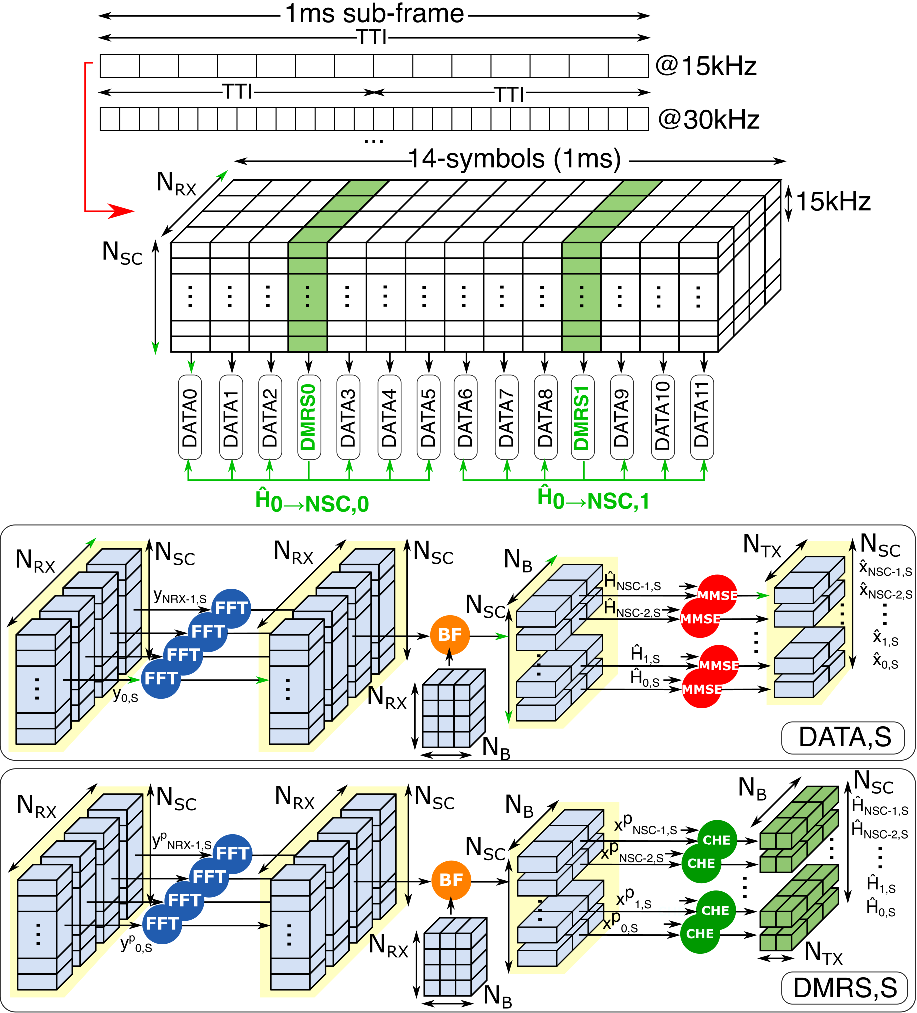}
  \caption{\textcolor{red}{Time-frequency grid of the samples received by the \gls{bs} in a \gls{pusch} TTI. The data processing for each symbol and the data dependencies between processing steps and symbols are highlighted in the dependency graph.}}
  \label{fig:pusch_pipeline}
\end{figure}

A \gls{5g} uplink transmission is organized in \si{10~ms}-frames, each containing \num{10} equal subframes. The subframe represented on top of \Cref{fig:pusch_pipeline} contains OFDM symbols, carrying information sent from \glspl{ue} to the \gls{bs} using \gls{ofdm}. \num{14} \gls{ofdm} symbols form a \gls{tti}. The scalable subcarrier spacing (from \si{15~kHz} to \si{960~kHz}) determines the \gls{tti} duration (from \si{1~ms} to \si{\num{0.015}~ms}). In this work we consider $\Delta f_{SC} = \si{15~kHz}$ subcarrier spacing (\si{1~ms} per \gls{tti}) and $N_{SC}=3276$ active subcarriers per \gls{ofdm} symbol~\cite{3GPP_TS_R17_description}. This results in the time-frequency grid on the top of \Cref{fig:pusch_pipeline}.

\begin{figure}[h]
  \centering
  \includegraphics[width=\columnwidth]{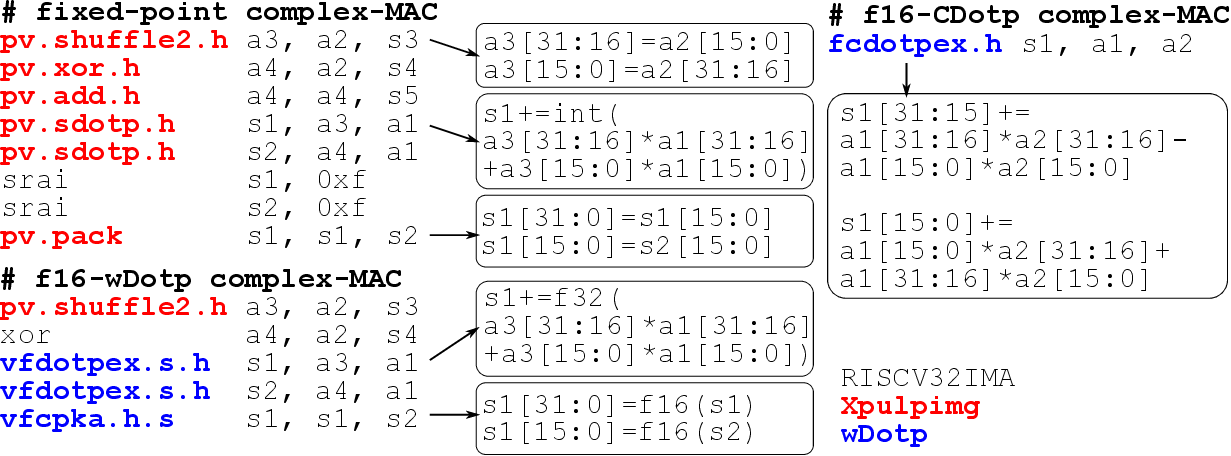}
  \caption{\textcolor{red}{Assembly code for complex MAC in different arithmetic precisions.}}
  \label{fig:complex_MAC}
  \vspace{-1em}
\end{figure}

The \gls{ofdm}-symbols are received by $N_{RX}$ antennas. The \gls{ofdm}-symbol processing at the \gls{gnb} is detailed in \Cref{fig:pusch_pipeline}. In our analysis \gls{pusch} carries data symbols and \glspl{dmrs}, the latter used to estimate the propagation channel. On the data symbols processing, we first encounter \gls{ofdm} demodulation, consisting of a per-antenna \gls{fft}. Second, \gls{bf}: antenna signals are linearly combined, generating $N_B$ beams, pointing to a specific transmitter. The key operator in \gls{bf} is a \gls{matmul} between the antenna streams and known coefficients. Third, an \gls{mmse} detector is applied for each subcarrier. In \gls{mimo} systems, $N_{TX}$ \glspl{ue} transmit on the same subcarrier. Each overlapping \gls{ue} per subcarrier is a \textit{layer}. The \gls{mmse} detector reconstructs the signal of each layer, minimizing the mean squared error between the transmitted and the reconstructed signal. 

Indicating with $k$ the subcarrier index, $s$ the symbol index, $y_{k,s}$ the $N_B$-long received symbol vector, $\hat{x}_{k,s}$ the $N_{TX}$-long reconstructed symbol, $\hat{H}_{k,s}$ the $N_{B}\times N_{TX}$ estimated channel coefficient matrix, ${\hat{H}_{k,s}}^H$ its Hermitian, $\sigma_{k,s}^2$ the channel noise variance and $I$ the identity matrix, the \gls{mmse} equation is:
\begin{equation}
    \hat{x}_{k,s} = ({\hat{H}_{k,s}}^H\hat{H}_{k,s} + \sigma_{k,s}^2I)^{-1}\hat{H}_{k,s}^Hy_{k,s}
    \label{eq:mmse}
\end{equation}

\textcolor{red}{The \gls{mmse} operator requires a matrix inversion. Given $G = ({\hat{H}_{k,s}}^H\hat{H}_{k,s} + \sigma_{k,s}^2I)$ and $b = \hat{H}_{k,s}^Hy_{k,s}$ from \Cref{eq:mmse}, we decompose $G=L^HL$ in its lower and upper triangular components, using Cholesky algorithm. The linear system $G\hat{x}=b$ can then be solved by inverting two lower triangular matrices $\hat{x}=L^{-1}\left((L^H)^{-1}b\right)$~\cite{Shahabuddin_MIMO_NorCAS_2020}.}

Since the transmission channel is not known a-priori to the receiver. ${\hat{H}_{k,s}}$ used in the formula is the result of the \gls{che} procedure that runs over the \num{2} \glspl{dmrs} in a \gls{tti}, with index $s = 0/1$. The data $x_{k,s}^p$ transmitted for \gls{dmrs} are \textit{pilots} known both at the transmitter and the receiver, which, given the received signal $y_{k,s}^p$, can therefore be used to estimate the channel through a classical \gls{lse} for the reference signals:

\begin{equation}
    \hat{h}_{ij,k,s} = \frac{{y^p}_{i,k,s}}{{x^p}_{j,k,s}}, \forall i \in [1, N_B], \forall j \in [1, N_{TX}] 
    \label{eq:che}
\end{equation}

\Cref{fig:pusch_pipeline} highlights the data dependencies between the \gls{pusch} processing steps. In this regard, \gls{dmrs}0/1 processing, in green, represents a critical block, as its output feeds the \gls{mmse} processing for the data symbols.

While the number of subcarriers can be selected to identify a high-load use case according to specifications, the number of antennas, beams, and layers are vendor-specific parameters. \gls{6g} requirements will push towards up to \num{128} antennas and up to \num{24} layers. Also, not all the slots in a frame are allocated to \gls{pusch}. Therefore, the following decoding operations for the \gls{pusch} symbol can stretch over the slot of other channels, while \si{1~ms} remains a hard limit for \gls{ofdm} computations, which are paced by the adopted \gls{tti}.

\subsection{Software-defined \gls{pusch} Operators Implementation}
\label{sec:PUSCH_op}

\textcolor{red}{We implement \gls{fp} and 16b fixed-point operators for \gls{5g}-\gls{pusch}. Inputs and outputs are complex (16b-real and 16b-imaginary). We describe the differences between implementations through the listings in \Cref{fig:complex_MAC}, that report the assembly for a complex \gls{mac} implemented with different integer and \gls{fp} \gls{isa} extensions, as an example. The \texttt{pv.shuffle.h}, \texttt{pv.pack}, and \texttt{vfdotpex.s.h} instructions do half-word shuffling and packing of 16b types in 32b types. The \texttt{pv.sdotp.h} and \texttt{vfdotpex.s.h} implement integer and \gls{fp} \gls{wdotp} of packed 16b-\gls{simd} types to 32b types. The fixed-point \gls{mac} always requires more instructions and registers than \gls{fp}: an arithmetic shift normalizes the result to fixed-point precision for each integer \gls{mac}. The \texttt{fcdotpex.h} is the most concise instruction, encoding the four \glspl{mac} required by the complex \gls{wdotp}.}

In our shared-memory cluster, the addresses of \gls{l1} follow word-level interleaving: they run contiguously over the 1st to the last Tile \gls{tcdm} 1st row, then switch to the 2nd memory row. The implementation of \gls{pusch} operators accounts for this: it profits from a fully-shared addressing space, minimizing the cores' conflict to shared memory resources, and reducing unnecessary out-of-Tile memory accesses.

\begin{figure}[h]
  \centering
  \includegraphics[width=\columnwidth]{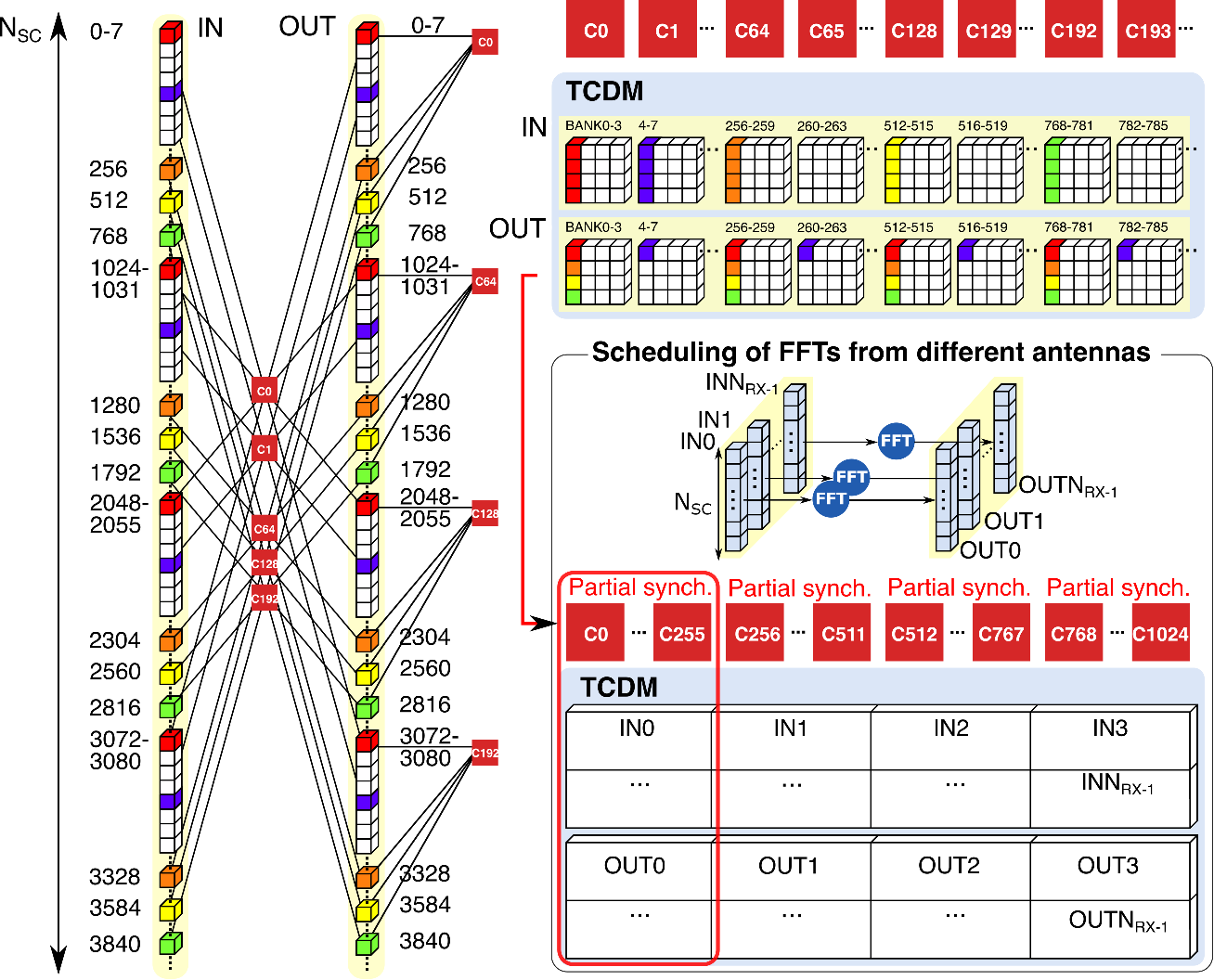}
  \caption{\textcolor{red}{Two butterfly stages of \gls{fft}, butterfly nodes are assigned to different cores. Inputs-outputs of a node have the same color. Top-right: cores fetch from local banks, and store inputs for the next stage. Bottom: parallelization of independent \glspl{fft} from different antennas.}}
  \label{fig:fft_algorithm}
\end{figure}

The \gls{fft} implementation follows Cooley-Turkey decimation in frequency, a multi-stage algorithm. In the butterfly stages of \Cref{fig:fft_algorithm}, input quartets produce output quartets (identified by the same color). The butterfly nodes are independent and we assign them to different cores for parallel processing. As shown in the top-right of \Cref{fig:fft_algorithm}, we store the input quartets in the local memory of the assigned processor, on different memory rows: they can be fetched with minimum latency. Outputs are stored to be fetched with minimum latency by the next \gls{fft} stage. With this approach, an \gls{fft} of $N$ complex samples lies on $N/4$ banks and is processed by $N/16$ cores.

\begin{figure}[h]
  \centering
  \includegraphics[width=\columnwidth]{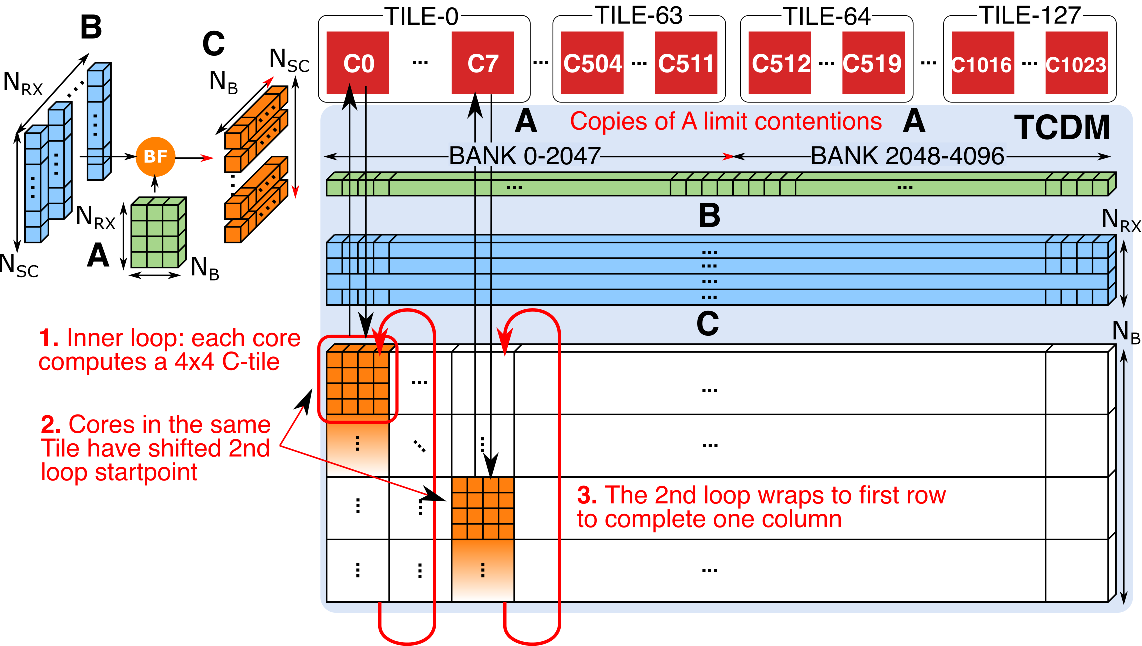}
  \caption{\textcolor{red}{\gls{matmul} parallelization. Cores are assigned the nearest copy of A in the addressing scheme and different columns of C.}}
  \label{fig:matmul_algorithm}
  \vspace{-1.5em}
\end{figure}

To fully utilize the cluster when $N/16<1024$, we assign inputs from different antennas to different core groups. For example, at the bottom-right of \Cref{fig:fft_algorithm} we execute four independent \glspl{fft} of \num{4096} points in parallel on groups of \num{256} cores. We separately synchronize cores working on independent \glspl{fft} with partial barriers. Synchronization is required after each butterfly stage. To reduce the overhead we let the cores work on the same stage of independent \glspl{fft} from different antennas in time sequence before the barrier.

\gls{bf} is a \num{3}-loop complex \gls{matmul} $C = A \times B$: matrix $A$ is a known $N_B \times N_{RX}$ coefficient matrix, $B$ is the $N_{RX} \times N_{SC}$ output of \gls{ofdm}. Parallelization is over the columns of $B$, on the subcarriers dimension. To maximize the compute intensity of the kernel, the loops are unrolled. Therefore, a single-core crosses multiple columns of $A$ and $B$, to produce an output tile in $C$. In the inner loop, all the inputs and partial sums must fit in the \gls{rf}, and the tile must be as big as possible, to maximize data reuse. A $2 \times 4$ output tile minimizes the variables pushed to the stack in the inner loop of the fixed-point and \gls{fp} \gls{wdotp} implementation. We use \num{16} registers for the real and imaginary parts of the accumulators, \num{4} registers for $A$ inputs, and \num{8} for $B$ inputs. In the complex-\gls{wdotp} \gls{fp} implementation, an output element is computed in a single Snitch instruction offload. This reduces the number of partial sum registers used. As a result, the output computation window grows to $4 \times 4$. Data in the inner loop is never pushed to the stack: we use \num{8} registers for inputs, \num{16} to accumulate temporary results, \num{3} for address increment, and \num{3} for loop control.

In our use case $A$, the $N_{B}\times N_{RX}$ \gls{bf} coefficient matrix, and $B$, the $N_{RX}\times N_{SC}$ data matrix from \gls{ofdm} are unbalanced in size. \Cref{fig:matmul_algorithm} shows how we distribute the data for this application in memory. Given the parallelization scheme, matrix $B$ is aligned to memory boundaries, reducing accesses outside a Tile. All the cores fetch elements in matrix $A$. To reduce the resulting congestion in the interconnect we adopt two solutions. First, we align matrix A to the memory rows and create multiple copies to fill up one row of the cluster's memory (in ~\Cref{fig:matmul_algorithm} we represent the case for $N_{B}\times N_{RX}=32\times64$, resulting in two copies of A). \textcolor{red}{When different beams apply to different groups of subcarriers, matrices with different coefficients can also be used, instead of A copies.} We then assign to each core the matrix A copy that is closer in the cluster addressing scheme. Second, we assign cores in the same Tile a different starting point in the loop across rows of A. This way it is less likely that cores will conflict for the same Tile port during out-of-Tile accesses. The bottom part of \Cref{fig:matmul_algorithm} represents different iterations of matrix C computation: the cores start iterating from different rows of C and go back to the first row, once they complete a column.

\gls{che} produces $N_{SC}$ matrices of size $N_{B}\times N_{TX}$, from $N_{SC}$ array couples of size $N_{TX}$ and $N_{B}$. Since the receiver holds prior knowledge of the \gls{dmrs} transmission, we assume that the kernel is fed the reciprocal of ${x^p}_{j,k,s}$ values in \Cref{eq:che}. In the parallelization, each core is assigned the generation of the channel estimate for a different subcarrier.

\begin{figure*}[ht]
  \centering
  \includegraphics[width=0.95\textwidth]{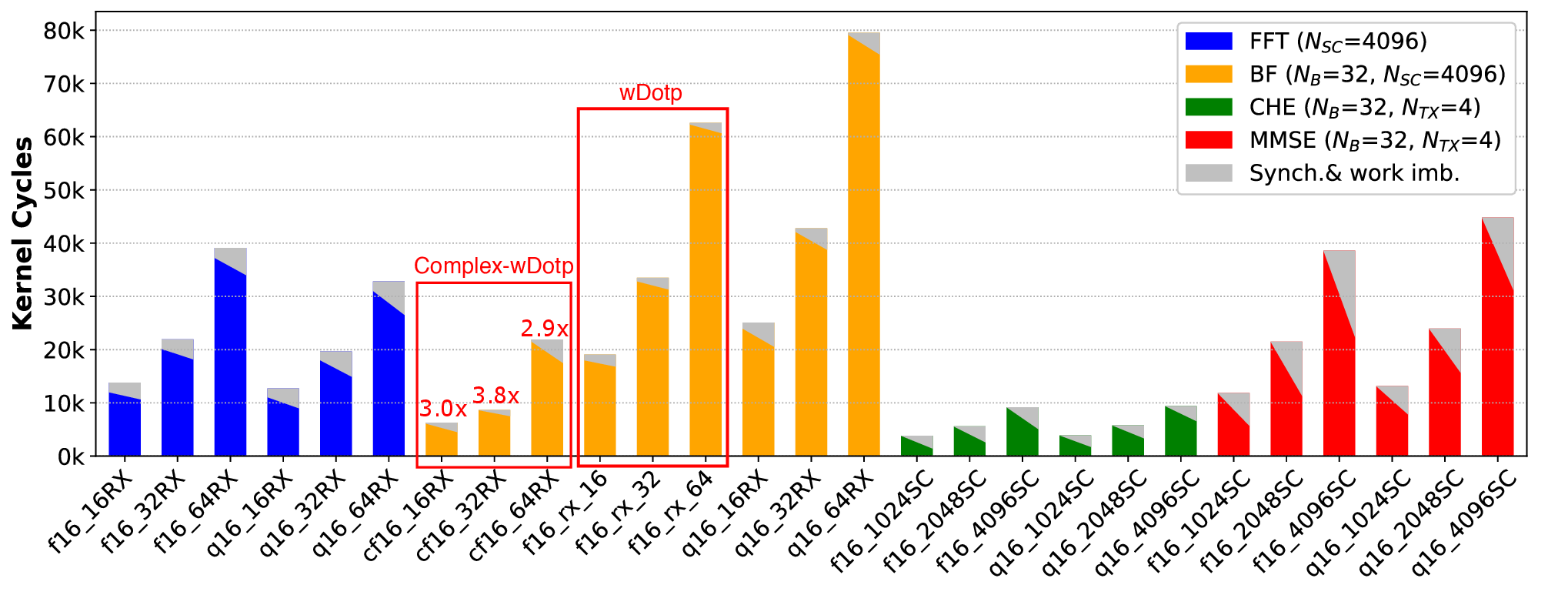}
  \caption{\textcolor{red}{Compute cycles of the implemented \gls{fp} and fixed-point kernels for different numbers of receiver antennas, subcarriers, beams, and layers. The skewed bars represent the arrival times of the first and last core to synchronization, the gray part represents synchronization and work-imbalance overheads.}}
  \label{fig:kernels_cycles}
  \vspace{-1em}
\end{figure*}
\begin{figure}[h!]
  \centering
  \includegraphics[width=\columnwidth]{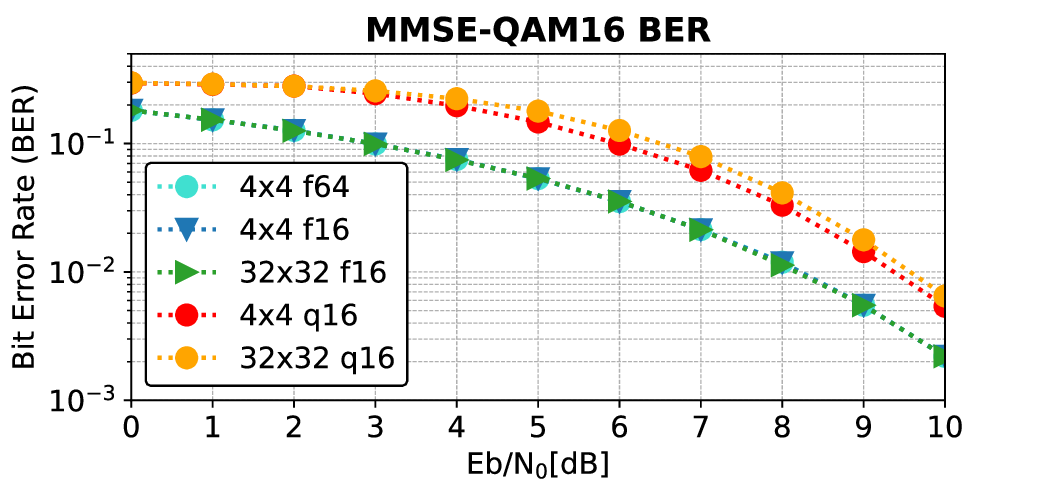}
  \caption{\textcolor{red}{\gls{ber} vs \gls{snr} of \gls{mmse} for different sizes and arithmetic precisions.}}
  \label{fig:ber}
  \vspace{-1em}
\end{figure}
\begin{figure*}[ht]
  \centering
  \includegraphics[width=\textwidth]{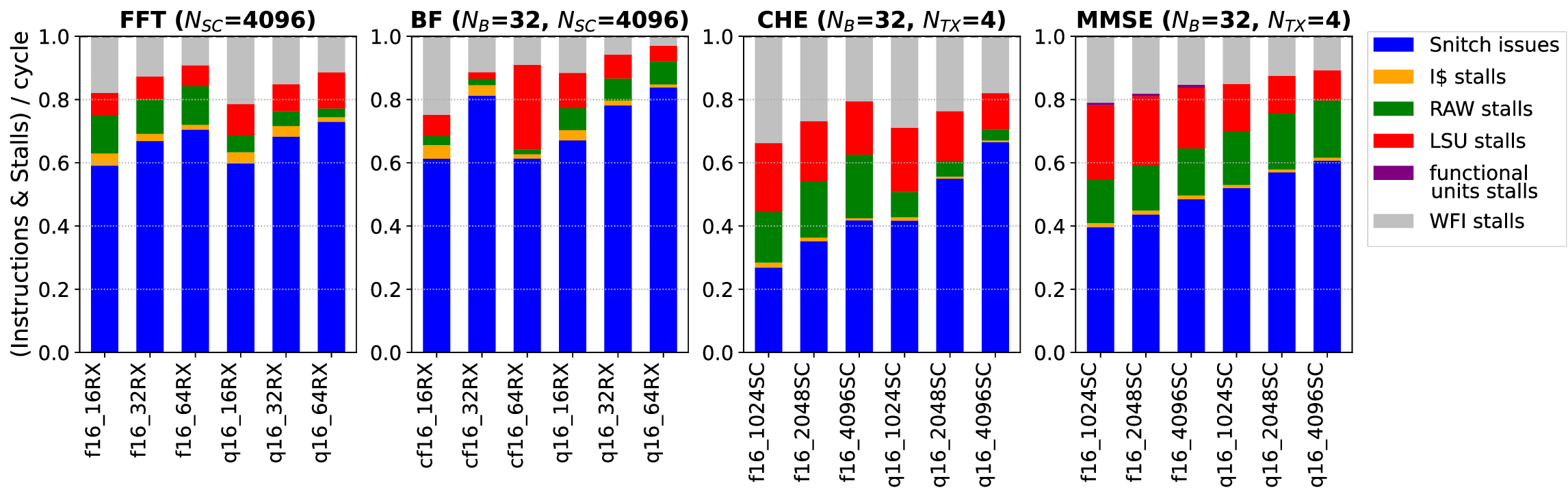}
  \caption{\gls{ipc} and architectural stalls breakdown of the implemented \textit{q16} and \textit{f16} kernels for different numbers of antennas, subcarriers, beams, and layers.}
  \label{fig:kernels_utilization}
  \vspace{-1em}
\end{figure*}
\begin{figure*}[ht]
  \centering
  \includegraphics[width=\textwidth]{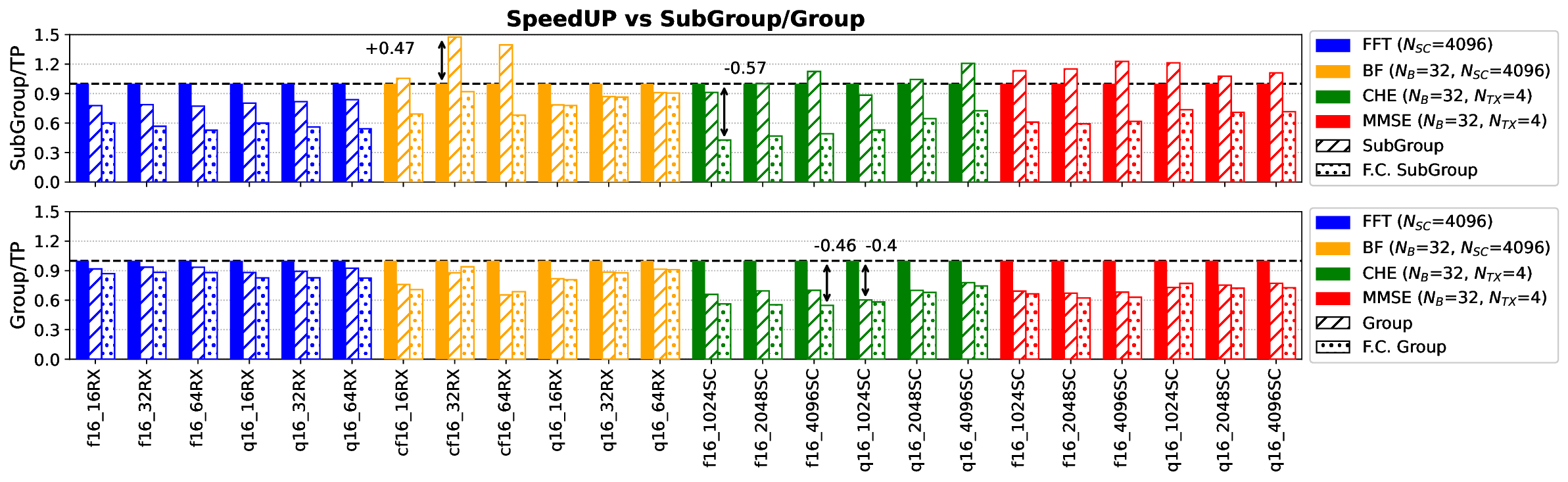}
  \caption{Ratio between the runtime of the full cluster and a SubGroup/Group on the implemented kernels. We consider both a SubGroup/Group with the same core-to-memory interconnect of TeraPool and an all-to-all fully connected SubGroup/Group. Input size is scaled over the parallelization dimension.}
  \label{fig:speedup}
  \vspace{-1em}
\end{figure*}

From \Cref{eq:mmse}, the \gls{mmse} operator requires the implementation of the channel Hermitian, the \gls{mvm} between the channel complex-conjugate and the received symbol, the Cholesky decomposition, and the solution of a triangular system. We implement these kernels for a single Snitch: the \gls{mmse} problem has strong inner data dependencies, harnessing efficient parallelization. Yet, the size of the inverted matrix depends on the $N_{TX}$, ranging between \num{4} and \num{32}, while $N_{SC}$ can be thousands, suggesting that the best choice is a coarse-grained parallelization of independent \gls{mmse} problems over the large dimensional subcarriers space.

\subsection{\gls{pusch} Operators Performance Analysis}
\label{sec:PUSCH_opperf}

\Cref{fig:kernels_cycles} represents the cycle-count of the implemented kernels \textcolor{red}{measured in \gls{rtl} cycle-accurate simulation of the cluster (Questasim 2022.3)}. We change the input size over the parallelization dimension of each kernel. For \gls{fft} and \gls{bf}, we set $N_B=32$ and $N_{SC}=4096$, and vary $N_{RX}$, for \gls{che} and \gls{mmse} we set $N_B=32$ and $N_{TX}=4$, and vary $N_{SC}$. \textcolor{red}{We mark the fixed-point and the \gls{fp} implementations with \textit{q16} and \textit{f16}. The most time-consuming processing step is \gls{bf} (\si{80~Kcycles/symbol} in \textit{q16} and \si{62~Kcycles/symbol} in \gls{fp}). Using \texttt{fcdotpex.h} (\textit{cf16}) we spare registers and instructions, reporting up to 3.8$\times$ speedup, compared to the cycle-count of \gls{fp}-\gls{bf} with \gls{wdotp} extensions only.} The \gls{fp} processing is bottlenecked by the \gls{fft} (\si{40~Kcycles} per symbol). \gls{che} is a small overhead: less than \si{10~Kcycles}, only required for \gls{dmrs} symbols. The skewed bars mark the arrival time of the first and last core to the barrier, clarifying how synchronization affects the runtime.

\textcolor{red}{In \Cref{fig:ber}, we also compare the arithmetic precision of the \textit{f16} against \textit{q16}, reporting the uncoded \gls{ber} of the \gls{mmse}, assuming a 16QAM modulation and transmission through additive white Gaussian noise channel. As expected~\cite{Ingemarsson_inversion_2015}, the \gls{ber} of \textit{q16} is higher than \textit{f16} and degrades on large problems.}

In \Cref{fig:kernels_utilization} we break down the average core instruction issues and architectural stalls over the total cycles, for each input dimension, and precision. Instruction stalls refer to cache misses, \gls{raw} stalls to cores waiting for data from the external pipelined and shared functional units or the memory interface. If the \gls{lsu} transaction table is full, the \gls{lsu} stalls the processor. The external functional units can also stall Snitch if their pipelines are busy. The \gls{wfi} stalls are encountered in synchronization when cores wait at the barrier for others to complete their job. The plots highlight \gls{lsu} stalls and \gls{raw} stalls as the main source of work imbalance. The \textit{f16}-\gls{bf} for $N_{RX}=64$ is an example: as replication of matrix A on the first memory row of interleaved banks is limited to two \num{32}$\times$\num{64} replicas, \gls{lsu} stalls and work imbalance increase. For \gls{che} and \gls{mmse}, the subcarrier-wise parallelization forces out-of-Tile accesses to data interleaved across all the banks. In turn, this increases \gls{lsu} stalls. We also observe more \gls{raw} stalls, caused by two main factors. First, in the subcarrier-wise parallelization, the loops are smaller. Data reuse is limited and more loads are required. This creates pressure on the interconnect and limits the throughput to the cores, now more often waiting for data. Second, in \gls{mmse} the cores must queue for division and square-roots executed in the shared unit. 

Nevertheless, working on large data chunks reduces synchronization overheads and increases the \gls{ipc}. An example is the \gls{fft}, which also benefits from the peculiar synchronization strategy: we compute a stage of the Cooley-Turkey algorithm for all the antennas and then synchronize. On average we achieve \num{0.8}-\num{0.68}~\gls{ipc} (\gls{fp} and fixed-point) on the per-antenna \gls{fft} and \gls{bf} parallelization, which takes most of the time in a \gls{tti}. We obtain \num{0.6}-\num{0.48}~\gls{ipc} on the more challenging \gls{che} and \gls{mmse} per-subcarrier parallelization. Considering the \gls{ipc} of the entire processing pipeline for the maximum dimensions under exam, the overall average \gls{ipc} is \num{0.75}-\num{0.61}.

To evaluate the effect of the hierarchical interconnect scale-up on parallelization, we run the kernels on a SubGroup/Group with respectively \num{64}/\num{256}-cores and \si{1~MiB}/\si{256~KiB} memory. 
We scale the input size over the parallelization dimension by the core-count ratio, ensuring that the workload fits in the SubGroup/Group memory, and Snitch executes the same number of instructions as in the full \num{1024}-core cluster.
\Cref{fig:speedup} shows the ratio between the operators' runtime on the full cluster and the SubGroup/Group. For \gls{bf} and \gls{mmse}, memory accesses spread over all the banks, and the full cluster performs better than the SubGroup: cores in a Tile will access different request-response ports, instead of competing for the same. The full cluster is at maximum \si{40~\%} (\gls{che}) slower than a Group. On average we measured \si{22~\%} slow-down, mainly caused by synchronization stalls, demonstrating a tolerable effect of the cluster hierarchical scale-up on parallelization.

We also consider an all-to-all fully connected SubGroup/Group. In this case, each core in a Tile has a memory request-response port to a 1~cycle latency \num{64}$\times$\num{64} or \num{256}$\times$\num{256} \gls{xbar}. Assuming these ideal yet physically unfeasible cluster configurations, our 1024-core shared-memory cluster obtains \si{35~\%}/\si{26~\%} average performance loss, compared to a SubGroup/Group. However, the biggest loss (up to \si{46~\%} and \si{57~\%}) is reported for \gls{che}, which requires less cycles than other benchmarks and does not influence the full chain performance.

Scaling up the shared memory hierarchically enables a physically feasible interconnect~\cite{Riedel_MemPool_2023, Zhang_TeraPool_2024}. Contentions for interconnect resources, bank conflicts, and synchronization overheads increase compared with the ideal all-to-all cluster. However, computing the same workload over smaller parallel clusters with private \glspl{tcdm} requires many copies of the shared data and forces solving data-dependencies between kernels executed on different clusters using producer-consumer queues and synchronizing across the clusters' hierarchies. This leads to data transfer overheads and increased program complexity. Multi-cluster parallelization is not desirable for \gls{pusch}, a large but rather uniform workload over the antennas (\gls{ofdm}) and the subcarriers (\gls{mmse}) dimensions, which can be easily parallelized on a single cluster.

\subsection{\gls{pusch} Operators Scheduling}
\label{sec:PUSCH_scheduling}

Alongside the processing pipeline building blocks, we also implemented full \gls{5g} applications, namely the sequence of \gls{ofdm} and \gls{bf}, and \gls{mimo}-\gls{mmse} detection. Combining the kernels for these applications, we consider the overhead of data movement, from the \gls{l1} scratchpad to the upper memory hierarchies, and we try to minimize them.

\gls{bf} combines the \gls{ofdm} antenna data streams to generate beam data streams. It must collect the output of $N_{RX}$ \glspl{fft}. The cluster \gls{l1} memory is large enough to fit the data for up to \num{64} antennas and can be used as a computing buffer. By stitching \gls{ofdm} and \gls{bf}, so that the data from \gls{fft} flows directly to the \gls{matmul} we avoid resorting to \gls{l2} transfers. When the \gls{bf} conditions do not change, executing multiple \gls{ofdm} symbols in sequence leads to the further advantage of keeping the coefficients matrix for \gls{bf} in \gls{l1}. We pay once the transfer from \gls{l2} and the replica of coefficients to the local memory of the cluster cores. Similarly, the output of \gls{che} is kept in \gls{l1}, to compute \gls{mmse} on adjacent symbols.

\begin{figure}[h]
  \centering
  \includegraphics[width=\columnwidth]{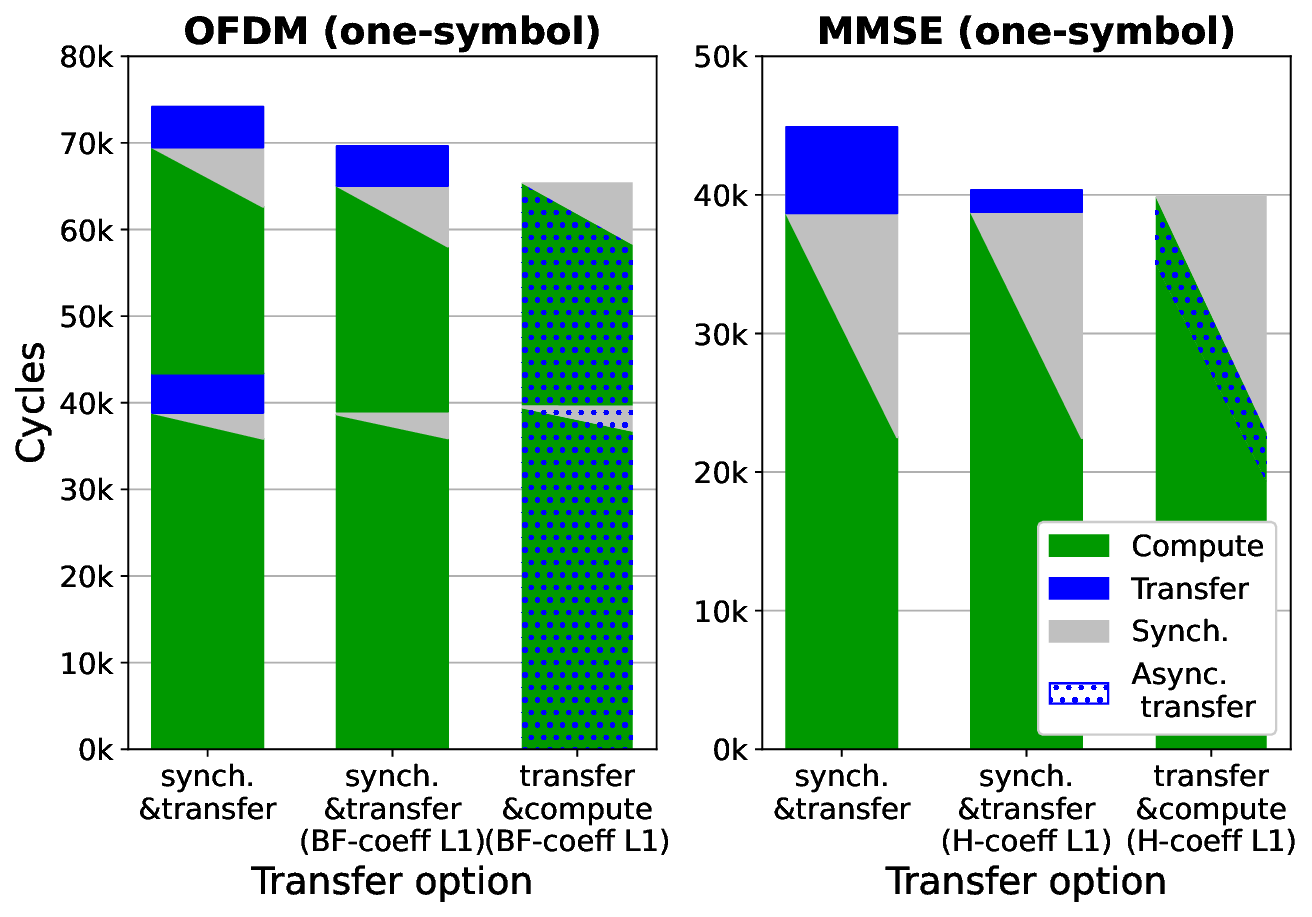}
  \caption{Execution cycles for an \gls{ofdm} and a \gls{mmse} symbol under different transfer options for inputs and outputs. The skewed bars represent the arrival of the first and last core to the synchronization barrier.}
  \label{fig:double_buffering}
\vspace{-1em}
\end{figure}
\begin{figure}[h]
  \centering
  \includegraphics[width=\columnwidth]{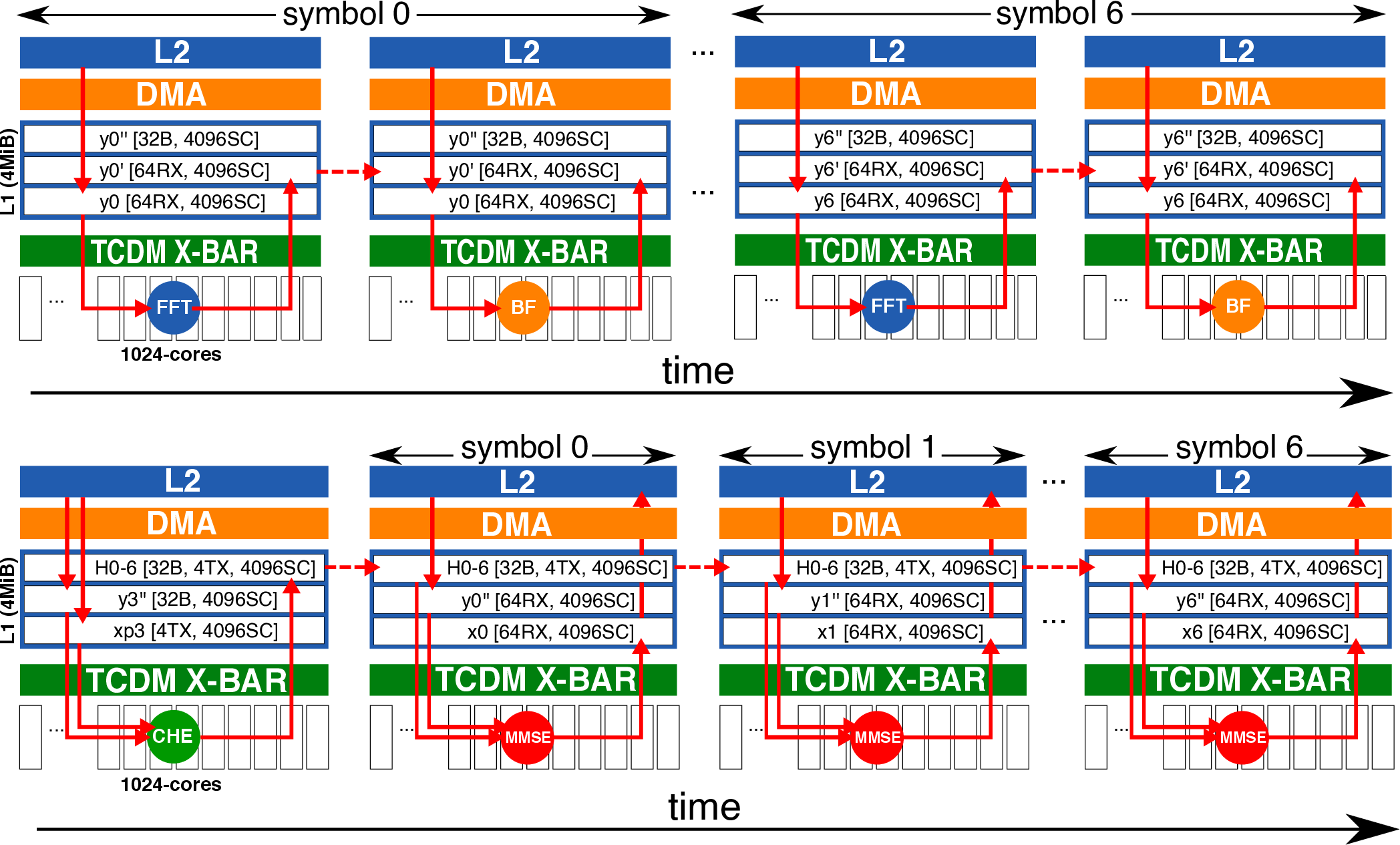}
  \caption{\textcolor{red}{Proposed execution pipeline for \num{7} symbols, in a system with $N_{RX}=64$, $N_B=32$, $N_{SC}=4096$, $N_{TX}=4$ layers per subcarrier. L1 data shared over different computation steps is marked by dashed arrows.}}
  \label{fig:pusch_scheduling}
\vspace{-0.5em}
\end{figure}

The benefit is demonstrated in \Cref{fig:double_buffering}, where we propose different strategies to transfer data for the computation of \gls{ofdm} and \gls{mmse} symbols. We try two approaches. In the first, the last core at the synchronization barrier transfers computed data and inputs for the next symbol. In the second, we instantiate a compute and a transfer buffer in memory. While computing, we transfer the data for the next symbol in the transfer buffer, using the \gls{dma}. During synchronization, we also wait for the \gls{dma}.  
The first bar in each plot of \Cref{fig:double_buffering} follows the first transfer method. It considers the transfer overhead of moving the \gls{bf} coefficients and the channel matrix from \gls{l2} at each symbol. In the case of \gls{ofdm} we also include the transfer of the \gls{fft} results to memory. Experiments with results reported in the second bars also follow the first transfer method. Using the large \gls{l1} as a compute buffer we reduce the total runtime of \si{6~\%} for \gls{ofdm} and of \si{11~\%} for \gls{mmse}. In the case of \gls{ofdm} we also benefit from directly pipelining \gls{fft} and \gls{bf} in \gls{l1}. The third bar shows the results for the second transfer method. The overlap between computation and asynchronous transfer by \gls{dma} engine is reported. During asynchronous transfers the \gls{dma} can conflict with cores for the scratchpad banks, affecting the runtime. For this reason, we start the \gls{mmse} transfers after the computation of the hermitian matrix, which requires heavy memory access. Double buffering gives an additional \si{6~\%} runtime improvement in \gls{ofdm}.

Finally, we propose the pipeline to execute a full \gls{pusch} \gls{tti}. We assume a \gls{gnb} system with $N_{RX}=64$, $N_B=32$, $N_{SC}=4096$, and $\Delta f_{SC} = \si{15~kHz}$. We consider $N_{TX}=4$ \glspl{ue} transmitting on each subcarrier, to test the processing chain under full load hypothesis. The execution pipeline for \num{7} \gls{tti} symbols is in~\Cref{fig:pusch_scheduling}, and it repeats for the next \num{7} symbols. Each replica of the cluster refers to a different computation step over time. The \gls{l1} data shared over multiple time steps is marked by dotted arrows, data transfers are represented by solid arrows. \gls{l1} memory is used as a low-latency buffer for the \gls{fft} output, feeding \gls{bf}, and the channel estimate, feeding \gls{mmse} over multiple symbols.

\section{Physical Implementation Results}
\label{sec:results}

In this section, we report the performance and area of our design, and the results of power simulations for key \gls{nr} kernels, running on the cluster. We then combine these data with execution time to perform an end-to-end performance and energy efficiency analysis for a \gls{pusch} \gls{tti} computation.

\subsection{Physical Implementation}

The physical feasibility of a baseline TeraPool was demonstrated in~\cite{Zhang_TeraPool_2024}. In this work, we challenged the implementation of the TeraPool cluster with Zfinx extensions and complex \gls{wdotp} in GlobalFoundries' 12~nm LPPLUS FinFET technology. We use Synopsys' Fusion Compiler 2022.03 for synthesis and \gls{pnr}. The Post-\gls{pnr} peak-performance \& area of our cluster, including \gls{fp} support are reported in~\Cref{tab:terapool_ppa}. 

\begin{table}[htb]
  \caption{Post-\gls{pnr} peak-performance \& area results.}
  \setlength{\tabcolsep}{2pt}
  \begin{threeparttable}
  \resizebox{\linewidth}{!}{%
    \begin{tabular}[h]{crcc}
      \toprule & & TeraPool & \textbf{This work} \\
      \multicolumn{2}{r}{Cluster Area [\si{\milli\meter\squared}]}                      & \num{68.9}    & \num{81.8}\\
      \multicolumn{2}{r}{SubGroup Area [\si{\milli\meter\squared}]}                     & \num{2.30}    & \num{3.01}\\
      \multicolumn{2}{r}{Area Per Core [\si{\milli\meter\squared/core}]}                & \num{0.067}   & \num{0.080}\\
      \multicolumn{2}{r}{Logic Gate Per Core [\si{MGE/core}]}                           & \num{0.17}    & \num{0.22}\\\midrule
      \multicolumn{2}{r}{Operating Frequency \emph{(Worst)} [\si{\mega\hertz}]}         & \num{637}     & \num{634}\\
      \multicolumn{2}{r}{Operating Frequency \emph{(Typ.)} [\si{\mega\hertz}]}          & \num{880}     & \num{875}\\
      \multicolumn{2}{r}{Peak Performance \emph{(Typ.)} [\si{TFLOPS}]}                         & \num{1.80}    & \num{1.79}\\
      \multicolumn{2}{r}{Peak Performance \emph{(Typ.)} [\si{Complex-TFLOPS}]}               & /    & \num{1.79}\\
      \midrule
    \end{tabular}}
  \end{threeparttable}
  \label{tab:terapool_ppa}
  \vspace{-0.5em}
\end{table}

Adding \gls{fp} support to the core complex improves the numerical precision of detection algorithms. Programming is also easier, as there is no need to rescale the representation range, depending on algorithms and operating conditions. The additional area cost is acceptable. The \gls{fp} unit is a significant overhead (+\si{40~\%}) on the core-complex area. On the Tile level, however, the area is \si{20~\%} of the total, and on the cluster level, it corresponds to an overhead of only \si{18~\%}. The routing channels introduced in the top-level of TeraPool by~\cite{Zhang_TeraPool_2024} do not change in size, resulting in no area waste for the cluster \gls{pnr}. Finally, the shared division and square-root units represent only \si{2~\%} of the Tile area, reducing the cost of these extensions, which would, on the contrary, be too expensive if added to the core (an additional estimated \si{20~\%}).

\begin{figure}[h]
  \centering
  \includegraphics[width=\columnwidth]{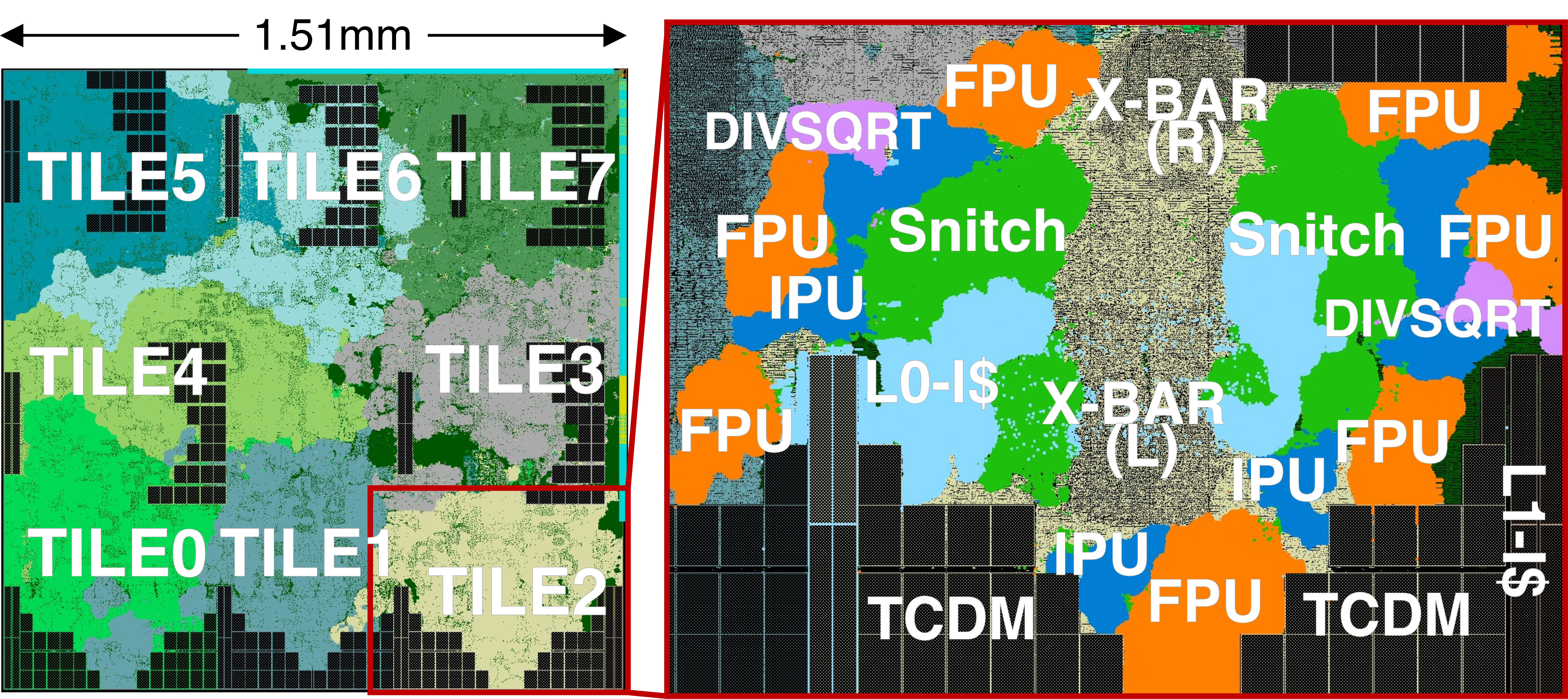}
  \caption{Snapshot of the SubGroup instance and zoom over a Tile, highlighting the FPUs, the IPUs, the Snitch cores, the division and square-root units, the \gls{icache}, and the \glspl{xbar}.}
  \label{fig:subgroup_snapshot}
  \vspace{-1em}
\end{figure}

The design does not incur operating frequency penalties compared to the cluster with cores only supporting integer instructions. It runs at \si{634~MHz} in the worst corner (\num{125}$^\circ C$, \si{\num{0.88}~V}), and at \si{875~MHz} in the typical corner (\num{25}$^\circ C$, \si{\num{0.8}~V}). At this frequency, the peak performance is \si{\num{1.79}~TOPS}, and the \textcolor{red}{L2 memory bandwidth (7.2Tbps) can sustain the raw \gls{iq} data-rate per antenna ($N_{RX}\times IQ_{bits}\times N_{SC}\times \Delta f_{SC} = 126Gbps$ in the target use-case with $IQ_{bits}=16b+16b$)}. \Cref{fig:subgroup_snapshot} represents a snapshot of the SubGroup instance and the Tile. The FPUs, the IPUs, the Snitch cores, the shared division and square-root units, the \gls{icache}, and the \glspl{xbar} are highlighted.

\begin{figure}[h]
  \centering
  \includegraphics[width=\columnwidth]{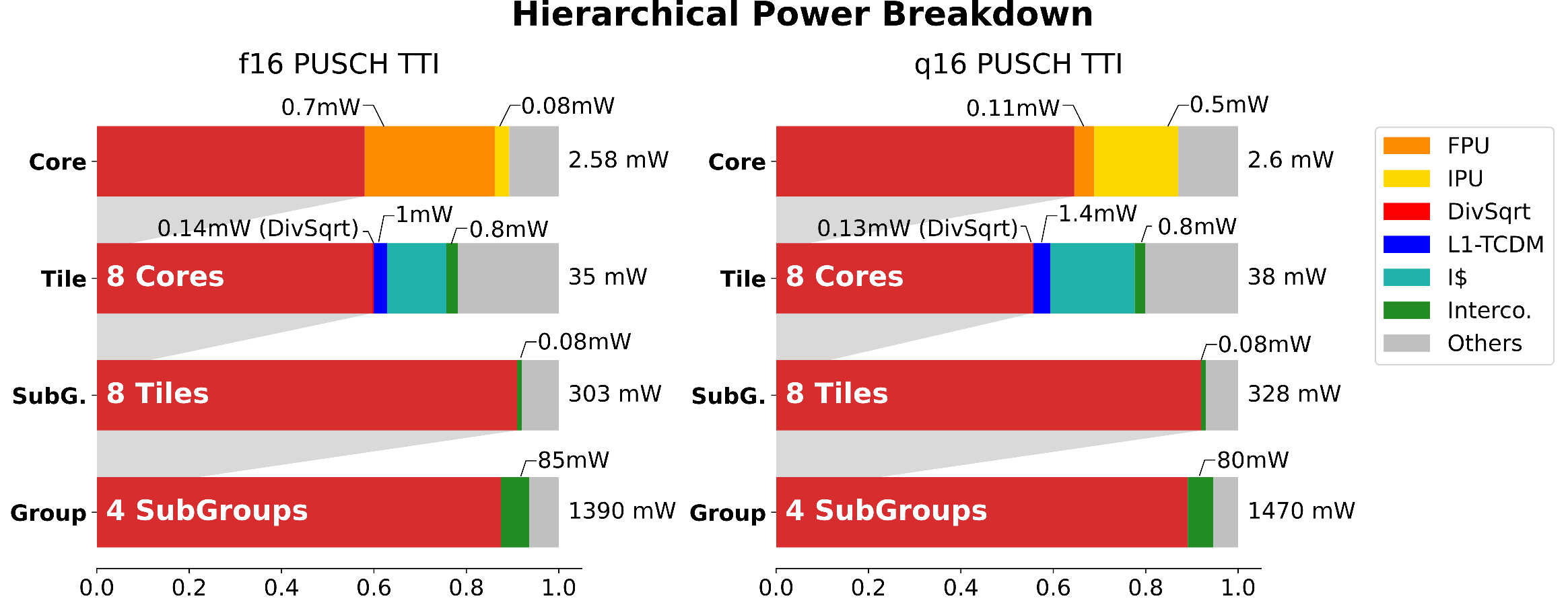}
  \caption{Breakdown over the hierarchies of the cluster of the power consumed to run a \gls{tti}. In \textit{others} we include the power of backend optimization cells and cells flattened by synthesis for timing optimization.}
  \label{fig:power_breakdown}
  \vspace{-1em}
\end{figure}

\begin{figure}[h]
  \centering
  \includegraphics[width=\columnwidth]{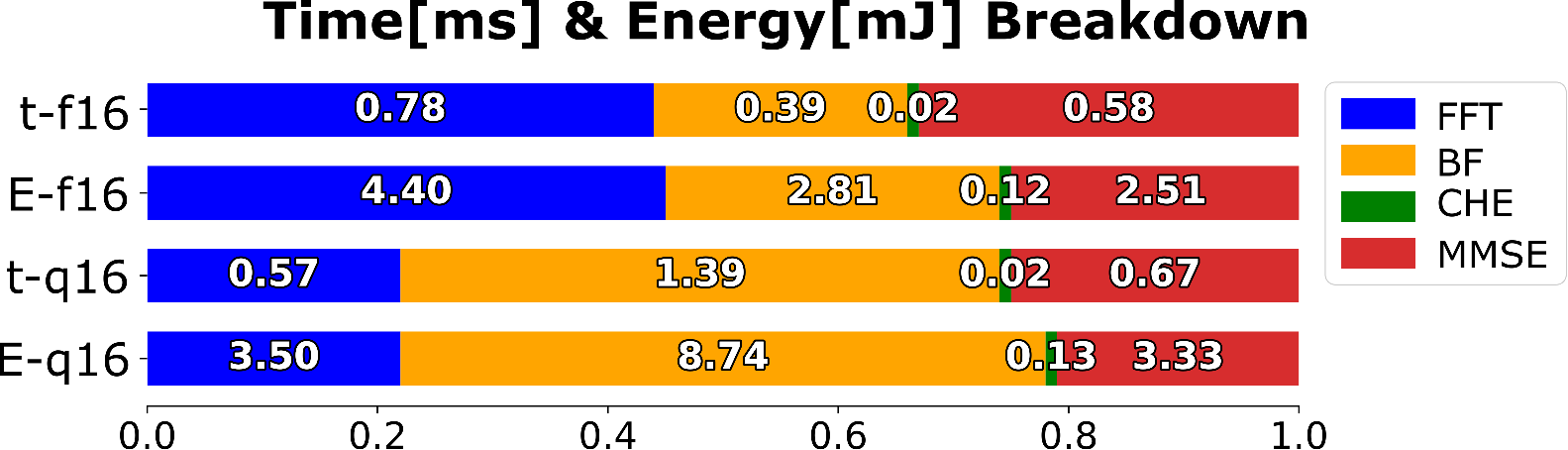}
  \caption{Breakdown of execution time and energy consumption over processing steps, for the \gls{pusch} computations in a \gls{tti}.}
  \label{fig:energy_breakdown}
  \vspace{-1em}
\end{figure}

\subsection{Throughput \& Power Simulations}

\textcolor{red}{
\Cref{tab:soa} reports the performance of the cluster on the kernels of \gls{pusch}. We consider an instance running in the typical corner (\num{25}$^\circ C$, \si{\num{0.8}~V}, at \si{800~MHz}), and the high load use case with $N_{SC}=4096$, $N_{RX}=64$, $N_{B}=32$, $N_{TX}=4$.  We compute the throughput metric depending on the processing step runtime, for a fair comparison with the \gls{soa}. For \gls{fft} and \gls{bf} (agnostic of $N_{TX}$) we consider the \gls{iq} antenna throughput: 
}
\begin{equation}
\frac{N_{RX} \times IQ_{bits} \times N_{SC}}{\#cycles} \times f_{ck}, \quad IQ_{bits} = (16b + 16b)
\label{eq:throughput1}
\end{equation}

\textcolor{red}{
\gls{che}, and \gls{mmse}, are agnostic of $N_{RX}$. We consider the raw \gls{iq} symbol throughput replacing $N_{RX}$ with $N_{TX}$ in \Cref{eq:throughput1}.
Finally, we report the throughput and power consumption for the \gls{pusch} processing in a \gls{tti} (\Cref{fig:pusch_pipeline}), executed as in \Cref{fig:pusch_scheduling}. We divide the antenna bits ($N_{RX}\times IQ_{bits} \times N_{SC} \times 14$) by the runtime measured in \gls{rtl} simulation (Questasim 2022.3). Power is extracted from post-layout simulations using Synopsys PrimeTime 2022-03.
}

The uplink achieves \si{44~Gbps/\num{5.9}W} for the \textit{q16} implementation (\gls{bf} is the bottleneck), and \si{\num{66.3}~Gbps/\num{5.54}W} for the \gls{fp} processing chain. \Cref{fig:power_breakdown} represents the breakdown of the main architectural components' power consumption across the cluster hierarchies. The Snitch cores and the functional units represent the main contribution to power consumption (\num{40}\%-\num{43}\%, fixed-floating). Accesses to scratchpad memories and the \gls{icache} contribute \num{0.18}\%-\num{0.12}\%. Finally, interconnections, including AXI, \gls{dma}, and \gls{tcdm} hierarchical \glspl{xbar}, contribute only \num{7.7}\%-\num{8.7}\%. This result highlights the energy efficiency of our hierarchical design. 

We estimate a total latency of \si{\num{2.6}~ms} (\textit{q16}) and \si{\num{1.7}~ms} (\textit{f16}). Running a \gls{tti} as in \Cref{fig:double_buffering} results in \si{60\mbox{-}100~Gbps} per \gls{ofdm} and \si{\num{7.7}\mbox{-}\num{8.8}~Gbps} per \gls{mmse} (\textit{q16}-\textit{f16}). \Cref{fig:energy_breakdown} shows the energy and time spent in each processing step for a \gls{tti}. The breakdown shows that for both the \textit{f16} and \textit{q16} implementations, most of the energy\&time is consumed by \gls{ofdm}. Running separately \gls{ofdm} and \gls{mmse}\&\gls{che} on two clusters, \gls{pusch} \gls{tti} throughput would be limited by the \gls{ofdm} stage, requiring \si{\num{1.96}~ms} in \textit{q16} and \si{\num{1.17}~ms} in \textit{f16}. The total average power consumption for the two clusters is \si{\num{11.8}~W} (\textit{q16}) and \si{\num{11.1}~W} (\textit{f16}).

\begin{table*}[h!]
  \centering
  \caption{Comparison of \gls{soa} hardware for lower-PHY 5G base station processing}
  \resizebox{\textwidth}{!}{
    \color{red}\begin{tabular}{lllllrrrr}
    \toprule
                & Technology & Area     & Frequency & Kernel & *\textsuperscript{\textdagger} Throughput & *Power & *Energy-Eff. & *Energy\&Area-Eff. \\
                &            & [mm$^2$] & & &[Gbps] & [W] & [Gbps/W] & [Gbps/W/mm$^2$]  \\
    \hline
    \hline
    \noalign{\vskip 1mm}
    \multirow{1}{*}{\textbf{This Work}}     &\multirow{1}{*}{GPP (GF-12nm)} &\multirow{1}{*}{81.8} &\multirow{1}{*}{800 MHz} &FFT q16 (64x4096)         &204.59	&6.09	&33.58	&0.41 \\
                                            &                               &                      &                         &FFT f16 (64x4096)         &150.38	&5.63	&26.70	&0.33 \\
                                            &                               &                      &                         &BF q16 (32x64x4096)       &84.40	&6.28	&13.44	&0.16 \\
                                            &                               &                      &                         &BF cf16 (32x64x4096)      &302.30	&7.22	&41.85	&0.51 \\
                                            &                               &                      &                         &CHE q16 (32x4x4096)       &43.59	&5.40	&8.07	&0.10 \\
                                            &                               &                      &                         &CHE f16 (32x4x4096)       &45.67	&4.56	&10.02	&0.12 \\
                                            &                               &                      &                         &MMSE q16 (16QAM 4x4x4096)       &9.36	&4.95	&1.89	&0.02 \\
                                            &                               &                      &                         &MMSE f16 (16QAM 4x4x4096)       &10.87	&4.34	&2.50	&0.03 \\
    \hline
                                            &                               &                      &                         &FFT\&BF q16         	    &59.75	&6.23	&9.60	&0.12 \\
                                            &                               &                      &                         &FFT\&BF f16         	    &100.43	&6.16	&16.30	&0.20 \\
                                            &                               &                      &                         &CHE\&MMSE q16       	    &7.71	&5.03	&1.53	&0.02 \\
                                            &                               &                      &                         &CHE\&MMSE f16       	    &8.78	&4.38	&2.00	&0.02 \\
                                            &                               &                      &                         &\textbf{PUSCH TTI q16} &\textbf{44.12}	&\textbf{5.90}	&\textbf{7.48}	&\textbf{0.09} \\
                                            &                               &                      &                         &\textbf{PUSCH TTI f16} &\textbf{66.3}	&\textbf{5.54}	&\textbf{11.96}	&\textbf{0.14} \\
    \noalign{\vskip 1mm}
    \hline
    \noalign{\vskip 1mm}
    \multirow{1}{*}{Amor \cite{amor_toc_2022}} &\multirow{1}{*}{GPP (GF-22nm \textit{synth.})} &\multirow{1}{*}{0.0081} &\multirow{1}{*}{100 MHz}   &FSK &0.15  &1.94  &0.08  &31.50 \\
                                               &                                               &                        &                           &LoRa &0.19 &1.96  &0.10  &41.18 \\
    \noalign{\vskip 1mm}
    \hline
    \noalign{\vskip 1mm}
    \multirow{1}{*}{Chen \cite{Chen_MIMO_TVLSI_2015}}                   &\multirow{1}{*}{ASIP (65nm)}      &\multirow{1}{*}{1.4}  &\multirow{1}{*}{400 MHz} &MMSE (64QAM 32x64)    &12.86	&0.02 &544.20	 &11405.02  \\
    \multirow{1}{*}{Kultala \cite{Kultala_Lord_TCAS_2019}}              &\multirow{1}{*}{ASIP (TSMC-28nm)} &\multirow{1}{*}{2.5}  &\multirow{1}{*}{968 MHz} &LORD (2x2)            &4.62	&0.09 &49.15	 &66.88\\
                                                                        &                                  &                      &                         &MMSE (64QAM, 4x4)     &6.25	&0.09 &66.54	 &90.54\\
    \multirow{1}{*}{Fu \cite{Fu_DREAM_TCAS_2022}}                       &\multirow{1}{*}{ASIP (TSMC-28nm)} &\multirow{1}{*}{4.35} &\multirow{1}{*}{500 MHz} &FFT (2048)            &52.73	&0.05 &1106.42   &1384.79\\
                                                                        &                                  &                      &                         &MMSE (256QAM, 128x8)  &15.03	&0.04 &335.93	 &420.45\\
    \multirow{1}{*}{Attari \cite{Attari_VectorASIP_TCAS_2022}}          &\multirow{1}{*}{ASIP (TSMC-28nm)} &\multirow{1}{*}{0.97} &\multirow{1}{*}{800 MHz} &ZF (64QAM, 128x8)     &6.75	&0.09 &79.25	 &272.92\\
                                                                        &                                  &                      &                         &MMSE (64QAM, 128x8)   &3.58	&0.09 &42.04	 &144.77\\
    \multirow{1}{*}{Castaneda \cite{Castaneda_MMSE_ASIC_ESSCIRC_2022}}  &\multirow{1}{*}{ASIP (GF-22nm)}   &\multirow{1}{*}{0.42} &\multirow{1}{*}{293 MHz} &MMSE (256QAM 128x16)  &1.76	&0.05 &33.26      &90.17\\
    \multirow{1}{*}{Chen \cite{Chen_DXT501_CoolChips_2022}}             &\multirow{1}{*}{ASIP (TSMC-28nm)} &\multirow{1}{*}{10.3} &\multirow{1}{*}{800 MHz} &FFT                   &7.31	&1.25 &5.83      &3.08\\
                                                                        &                                  &                      &                         &CHE                   &7.31	&1.25 &5.83      &3.08\\
                                                                        &                                  &                      &                         &MIMO (16QAM 2x2)      &7.31	&1.25 &5.83      &3.08\\
    \noalign{\vskip 1mm}
    \hline
    \noalign{\vskip 1mm}
    \multirow{1}{*}{Peng \cite{Peng_MMSE_ASIC_TCAS_2018}}           &\multirow{1}{*}{ASIC (28nm)}                      &\multirow{1}{*}{2.57} &\multirow{1}{*}{680 MHz} &MMSE (64QAM 128x8)  &29.47	 &0.08	&383.68	 &4380.27 \\
    \multirow{1}{*}{Tang \cite{Tang_MMSE_ASIC_JSSCC_2021}}          &\multirow{1}{*}{ASIC (40nm)}                      &\multirow{1}{*}{0.58} &\multirow{1}{*}{425 MHz} &MPD (256QAM 128x32) &27.60	 &0.05	&529.26	 &10139.11 \\
    \multirow{1}{*}{Shahabuddin \cite{Shahabuddin_MIMO_TVLSI_2021}} &\multirow{1}{*}{ASIC (TSMC-65nm)}                 &\multirow{1}{*}{0.69} &\multirow{1}{*}{606 MHz} &ADMIN (64QAM 64x32) &4.79	 &0.04	&129.31	 &1023.25 \\
    \multirow{1}{*}{Guo \cite{Guo_FFT_ASIC_TVLSI_2023}}             &\multirow{1}{*}{ASIC (TSMC-12nm)}                 &\multirow{1}{*}{0.75} &\multirow{1}{*}{1 GHz}   &FFT (4096)          &99.20	 &0.78	&127.48	 &169.97 \\
    \multirow{1}{*}{Yang \cite{Yang_FFT_ASIC_TCAS_2023}}            &\multirow{1}{*}{ASIC (SMIC-40nm \textit{synth.})} &\multirow{1}{*}{0.84} &\multirow{1}{*}{483 MHz} &FFT (2048)          &69.76	 &-		&-		 &- \\
    \multirow{1}{*}{Zhang \cite{Zhang_MMSE_ASIC_ISSCC_2024}}        &\multirow{1}{*}{ASIC (40nm)}                      &\multirow{1}{*}{4.36} &\multirow{1}{*}{200 MHz} &BP (16QAM 8x8)      &32.77  &0.07	&452.85  &1154.04 \\
    \noalign{\vskip 3mm} 
    \end{tabular}
  }
  \label{tab:soa}
  \small\raggedright\textsuperscript{*}Throughput, power, and area were normalized to GlobalFoundries' 12nm LPPLUS FinFET technology, multiplying respectively by the feature size ratio $s$, $\frac{1}{s}(\frac{0.8V}{VDD})^2$, $\frac{1}{s^2}$.\\
  \small\raggedright\textsuperscript{\textdagger}Raw \gls{iq}-data throughput. The results for \glspl{asip} and \glspl{asic} are computed according to adopted data-width and modulation scheme.
\end{table*}

\begin{table*}[h!]
  \centering
  \caption{Industry SoCs for lower-PHY 5G base station processing}
    \begin{tabular}{lllllc}
    \toprule
        Platform & PHY processors & ISA & Multi-core & SW-defined operators & 5G-split \\
    \hline
    \hline
    \multirow{1}{*}{NVIDIA AX800 \cite{ax800}}     &Ampere GPU                 &NVIDIA &yes (8192c) &cuMAC              &8\\
                                                    &BlueField-3 DPU            &       &            &cuPHY              &\\
    \hline
    \multirow{1}{*}{EdgeQ/S-series \cite{EdgeQ}}    &TXU processor              &RISC-V &yes (60c) &Demodulation              &6-7.2\\
                                                    &ARM Neoverse-E1            &ARM   &yes (8c)  &Beamforming               &\\
                                                    &                           &      &         &64x64 MIMO                &\\
                                                    &                           &      &         &Matrix decomposition      &\\
                                                    &                           &      &         &Channel Estimation        &\\
                                                    &                           &      &         &Equalization              &\\
    \hline
    \multirow{1}{*}{Picocom/PC802 \cite{Picocom_PC802}}    &Ceva XC12 1280-bit           &RISC-V &no        &Digital Front-End      &7.2X\\
                                                           &Scalar-processors cluster    &RISC-V &yes (25c) &Encoding-Decoding      &\\
                                                           &                             &      &          &Demodulation           &\\
                                                           &                             &      &          &4x8 MIMO               &\\
    \hline
    \multirow{1}{*}{Marvell/Octeon10-CNF105xx \cite{Marvell_Octeon10}}     &ARM Neoverse-N2  &ARM   &yes ($<$36c) &64x64 MIMO     &7.X\\
                                                                           &DSP processors   &n.a.  &yes        &Others (n.a.)  &\\
                                                                           &Accelerators     &      &           &               &\\
                                                                           &                 &      &           &               &\\
    \hline
    \multirow{1}{*}{Qualcomm/X100 \cite{Quallcom_X100}}   &n.a.  &n.a. &n.a. &Demodulation     &7.X\\
                                     &      &     &     &Beamforming      &\\
                                     &      &     &     &64x64 MIMO       &\\
                                     &      &     &     &Channel Coding   &\\
    \hline
    Platform & PHY processors & \multicolumn{3}{c}{HW-defined operator} & 5G-split \\
    \hline
    \hline
    \multirow{1}{*}{Xilinx/Zynq-Ultrascale\&RFSoC \cite{Xilinx_RFSoC}}      &Programmable Logic    &\multicolumn{3}{c}{Demodulation} &7.X\\
                                                                            &ARM Cortex-A53 (4c)   &\multicolumn{3}{c}{PRACH} &\\
                                                                            &Accelerators          &                          & & &\\
    \end{tabular}
  \label{tab:industry}
  \vspace{3em}
\end{table*}

\section{Related Works}
\label{sec:related_works}

This section compares our cluster with \gls{soa} solutions for \gls{gnb} processing. We group and compare the processing engines for \gls{5g}-\gls{nr} from academic literature in three main categories, based on their programmability (\Cref{tab:terapool_ppa}). Under \gls{gpp}~\cite{amor_toc_2022}, we report \gls{dsp} solutions based on programmable processors, supporting a full \gls{isa}. These are the most flexible in programmability and the best candidates to pursue software-defined wireless processing. 

In the \gls{asip} category we report programmable devices, with a reduced custom \gls{isa} tailored to the \gls{gnb} workload~\cite{Chen_MIMO_TVLSI_2015, Kultala_Lord_TCAS_2019, Fu_DREAM_TCAS_2022, Attari_VectorASIP_TCAS_2022, Castaneda_MMSE_ASIC_ESSCIRC_2022, Chen_DXT501_CoolChips_2022}. Compared to \glspl{gpp}, they target higher frequencies at lower power consumption, but they require architecture specialization. As a result \glspl{asip} will execute a limited number of workloads, with fixed problem sizes. The \gls{vliw} architecture chosen by \cite{Fu_DREAM_TCAS_2022} and \cite{Attari_VectorASIP_TCAS_2022} to maximize task specialization also adds complexity to the programming model, exacerbating the issue. This is not optimal for software-defined \gls{gnb} processing, considering the fast-paced evolution of telecommunication standards, input data dimensions, and algorithms. 

The \gls{asic} category includes accelerators tailored to specific workloads, targeting high performance, with extreme architecture specialization~\cite{Peng_MMSE_ASIC_TCAS_2018, Tang_MMSE_ASIC_JSSCC_2021, Shahabuddin_MIMO_TVLSI_2021, Guo_FFT_ASIC_TVLSI_2023, Yang_FFT_ASIC_TCAS_2023}. These engines have limited reconfigurability options and can be used to accelerate \gls{gnb} workloads in heterogeneous systems, with programmable cores handling data movement and control operations.

\begin{figure}[ht]
  \centering
  \includegraphics[width=0.95\columnwidth]{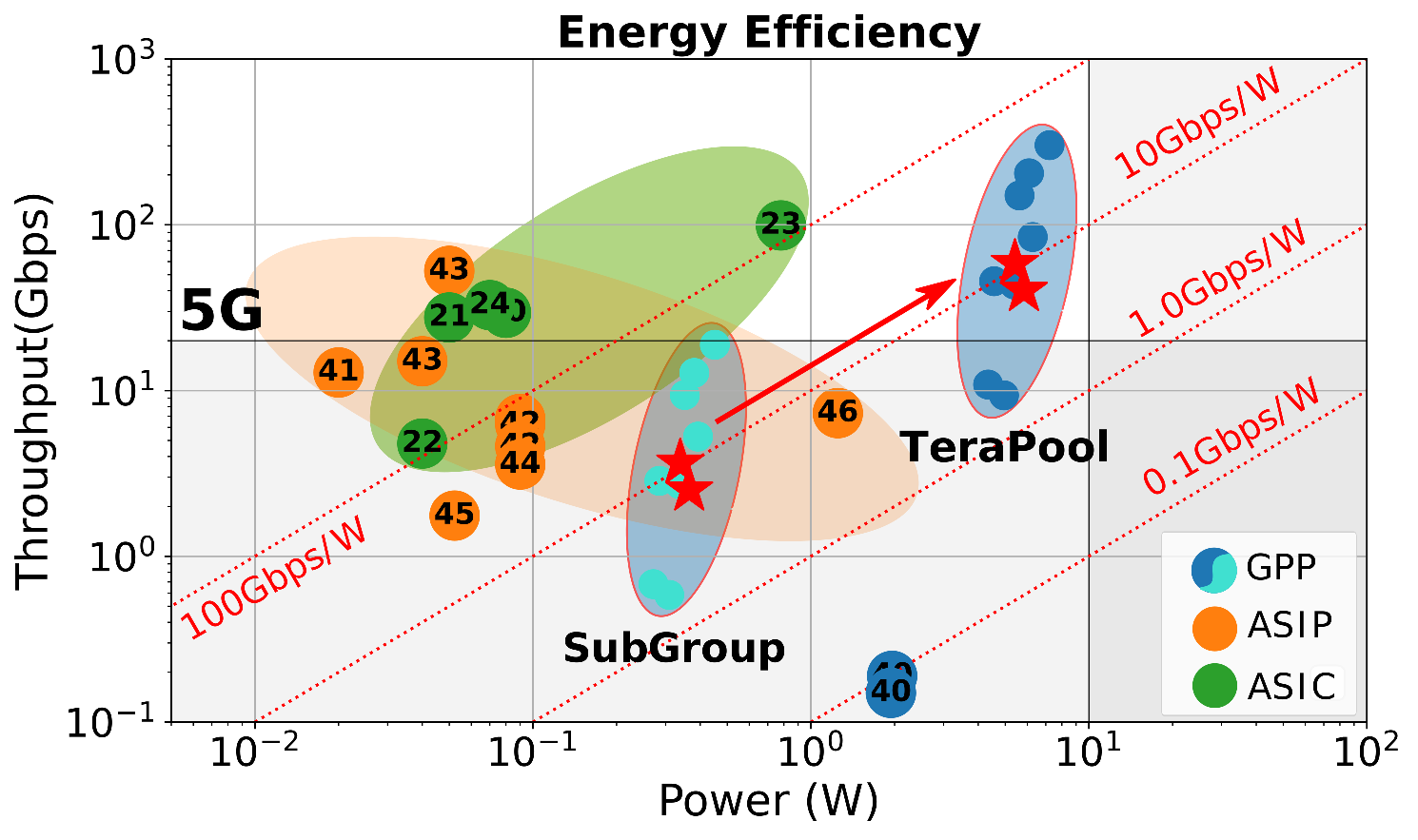}
  \caption{Comparison of the energy efficiency of our cluster with ASIC and ASIP \gls{soa} solutions for \gls{gnb} processing. The star represents the results for one \gls{pusch} symbol.}
  \label{fig:energy_efficiency}
  \vspace{-1em}
\end{figure}

Our solution is a cluster of \gls{isa}-enhanced RISC-V cores. Its position in the \gls{soa}, compared to \glspl{asip} and \glspl{asic} is clarified in~\Cref{fig:energy_efficiency}, where the focus is on energy efficiency. To allow a fair comparison, the throughput, power, and area of the proposed \gls{soa} solutions were normalized to GlobalFoundries' \si{12~nm} LPPLUS FinFET technology\footnote{Throughput, power, and area where normalized multiplying respectively by the feature size ratio $s$, $\frac{1}{s}(\frac{0.8V}{VDD})^2$, $\frac{1}{s^2}$}. Our many-core has \num{1024} cores running at high frequency, thanks to an aggressive backend design in an advanced technology node. Its \gls{fft}-throughput largely outperforms the \gls{soa} single-core \gls{gpp} solution ($>1000\times$ on a similar workload). TeraPool achieves up to $6.2\times$ (\gls{mmse}), and \si{28\times} (\gls{fft}), higher throughput than \glspl{asip}, and up to \si{2\times} higher \gls{fft}-throughput than the \glspl{asic} in our survey. Differently from the \gls{soa} \glspl{asip} and \glspl{asic}, the programmable TeraPool cores support a full uplink workload. Its \si{\num{66.3}Gbps} \gls{tti}-throughput fulfills the \gls{5g} \si{20Gbps} uplink requirement and follows the increasing rates of \gls{6g} workloads~\cite{ITU_2017, Rappaport_THz_IEEE_2019}. The penalty is higher power consumption. The programmable cluster requires up to \si{100\times} more average power than specialized datapaths to execute \gls{6g} kernels. In ~\Cref{fig:energy_efficiency} we report the throughput provided by a SubGroup versus its power consumption in kernels execution.

\textcolor{red}{
In comparing energy efficiency with ASIPs and ASICs, it is important to note that all  published results target a much lower throughput and are measured on only a subsed of a functionally complete PUSH workload.  Clearly, the functional flexibility and high throughput  of our cluster have a price in area and energy efficiency.  To put this statement in perspective, let's  consider \cite{Kultala_Lord_TCAS_2019}, which has \num{128} complex \gls{mac} units, the highest in our survey. A SubGroup (\num{64} cores) achieves similar energy and area efficiency. The area efficiency reduction resulting from the scaling up of our cluster to match beyond \gls{5g} and \gls{6g} requirements is mainly caused by the top-level routing channels, as highlighted by~\cite{Zhang_TeraPool_2024}. However, a bigger shared-memory cluster offers two main advantages. First, the large \gls{l1} can entirely accommodate the data footprint of a \gls{pusch} workload. As demonstrated, it can be used as compute buffer, to hide transfer latencies. It also simplifies the programming model: we benefit from massive \gls{spmd}, with no need for data transfers and synchronization between many small clusters. Second, the cluster is more versatile than extremely specialized programmable solutions, such as \cite{Fu_DREAM_TCAS_2022} and \cite{Attari_VectorASIP_TCAS_2022}, resulting in a better match for software-defined lower-\gls{phy} workloads.
}

\textcolor{red}{
Finally, considering \si{10~W} power budget per \gls{bs} \gls{dsp} component~\cite{Huawei_Power, Blume_IEEE_2010}, our scaled-up solution matches the required power consumption and exceeds by $6\times$ the energy efficiency constraints (\si{2~Gbps/W}). We observe that energy efficiency is constant if we consider the scaling of one SubGroup to the full cluster, for minimal energy overhead of interconnections.}

We also compare our approach to rapidly evolving industry solutions for software-defined \gls{5g} processing. Relevant examples are reported in ~\Cref{tab:industry}. \glspl{gpu} achieve high-performance parallelization and keep software versatility. They are nowadays the main computing platforms for deep-learning applications, and they are also considered good options to fill the flexibility/performance gap required by \gls{5g} and beyond~\cite{tarver_gpu_2021}.
In fact, aiming for the reuse of existing \gls{gpu} cloud infrastructure in the domain of \gls{5g}, NVIDIA fostered \gls{ran} virtualization~\cite{kundu_gpu_2023}. Wireless processing software libraries~\cite{kelkar_aerial_2021, hoydis_sionna_2022} with CUDA support were developed to run on commodity cloud hardware, such as the AX800 converged accelerator~\cite{ax800}, which achieves \si{\num{36.56}~Gbps} (downlink) and \si{\num{4.79}~Gbps} (uplink) throughput. The performance of \glspl{gpu} on wireless workloads was also revealed by academic research: \cite{tarver_gpu_2021} achieves \si{\num{3.96}~Gbps} on LDPC decoding, previous works on older \glspl{gpu} generations report \si{\num{0.13}\mbox{-}\num{0.52}~Gbps} on parallel decoding of many independent \gls{mimo} systems~\cite{wu_gpu_2014, li_gpu_2015}. An analysis of the power consumption of these platforms on \gls{5g} benchmarks is however still not disclosed in the open literature. With \si{350~W} peak power consumption, AX800 usage is recommended in a cloud infrastructure~\cite{ax800}. \gls{gpu} based \gls{ran} processing, therefore targets \gls{5g} split 8~\cite{Larsen_COMST_2019}, where none of the \gls{phy} functions is implemented in the \gls{bs}, putting high requirements on the data rates of the network fronthaul: \si{100~Gbps}~\cite{kundu_gpu_2023}.

\textcolor{red}{
The \gls{fpga} solutions~\cite{Xilinx_RFSoC} accelerate the computationally intensive uplink operations on the programmable logic, while data transfers and control are executed on a multi-core \gls{cpu}. Despite lower time-to-market than \glspl{asic}, adapting the \gls{gnb} to programmable logic requires more implementation and verification effort, compared to a streamlined software approach.}

\cite{EdgeQ, Picocom_PC802, Marvell_Octeon10} are heterogeneous platforms all containing one or more programmable components. They implement the \gls{5g} split 7.X, where the \gls{phy} functions up to decoding of the received signals are offloaded to the \gls{bs}~\cite{Larsen_COMST_2019}. For all these platforms, vendors offer pre-built libraries of kernels with the hardware.

Our approach is aligned with this trend: it can be used to implement a software defined version of split 7.X and it offers additional advantages. First, it has at least \si{10\times} more cores than all the clusters used in industry platforms for deployment at the network's edge. Its processing elements are however smaller and architecturally simple. The core complex occupies only \si{80~kGE} and a SubGroup instance, including \num{64} cores and \si{1~MiB} of memory occupies \si{\num{2.3}~mm^2} in \si{12~nm}. The ARM Neoverse-N2 in~\cite{Marvell_Octeon10} is more bulky: a single core occupies \si{\num{1.3}~mm^2}~\cite{ARM_NeoverseN2} in \si{5~nm}. Second, it offers competitive power consumption. It consumes \si{\num{5.54}~W} on average for \gls{pusch} uplink. This processing is similar to the downlink workload described by~\cite{Quallcom_X100}, which in turn consumes \si{\sim18~W} on the X100 card. Third, it proposes an efficient novel shared-memory architecture. This work showed how this can be used to implement an in-line processing pipeline, where the \gls{l1} memory is used as a low-latency compute buffer.

\section{Conclusion}
\label{sec:conclusions}

This paper addressed the three main problems arising from the scale-up of beyond \gls{5g} transmission bandwidth and number of subcarriers: throughput, energy efficiency, and lifetime of \gls{bs} hardware components. The proposed solution is a many-core cluster of \num{1024} lightweight RISC-V processing elements, sharing \si{4~MiB} of \gls{l1} memory. The high core count pushes peak performance on key \gls{nr} kernels to \si{20~Gbps}, beyond \gls{5g} requirements. We estimate average power consumption of \si{<6~W} and a total latency of \si{\num{2.6}~ms} (\textit{q16}) and \si{\num{1.7}~ms} (\textit{f16}) for the execution of \gls{pusch}. The big shared memory reduces expensive transfers of large \gls{nr} data, maximizing the computing phases of processing. Scaling up the shared-memory cluster, using hierarchical \gls{xbar} interconnects keeps energy efficiency high. Our design achieves \si{\num{9.4}\mbox{-}300~Gbps/W} on all the kernels under test, up to \si{28\times} larger than other \glspl{asip} in the \gls{soa}. It executes a \gls{pusch} symbol at \si{12~Gbps/W}. Finally, the programmable RISC-V Snitch core and the cluster' streamlined programming model enable a full software-defined processing approach, a key asset to keep up with evolving telecommunication standards.

\section*{Acknowledgment}
This work was funded by Huawei Sweden AG, and by the COREnext project supported by the EU Horizon Europe research and innovation program under grant agreement No. 101092598.

\bibliographystyle{IEEEtran} 
\bibliography{REFERENCES.bib}

\begin{thebibliography}{10}
\providecommand{\url}[1]{#1}
\csname url@samestyle\endcsname
\providecommand{\newblock}{\relax}
\providecommand{\bibinfo}[2]{#2}
\providecommand{\BIBentrySTDinterwordspacing}{\spaceskip=0pt\relax}
\providecommand{\BIBentryALTinterwordstretchfactor}{4}
\providecommand{\BIBentryALTinterwordspacing}{\spaceskip=\fontdimen2\font plus
\BIBentryALTinterwordstretchfactor\fontdimen3\font minus \fontdimen4\font\relax}
\providecommand{\BIBforeignlanguage}[2]{{%
\expandafter\ifx\csname l@#1\endcsname\relax
\typeout{** WARNING: IEEEtran.bst: No hyphenation pattern has been}%
\typeout{** loaded for the language `#1'. Using the pattern for}%
\typeout{** the default language instead.}%
\else
\language=\csname l@#1\endcsname
\fi
#2}}
\providecommand{\BIBdecl}{\relax}
\BIBdecl

\bibitem{3GPP_TS_R15}
{3GPP}, ``{Release 15 Description; Summary of Rel-15 Work Items},'' {3rd Generation Partnership Project (3GPP)}, Technical Specification (TS) 21.915, 2017, {Release 15}.

\bibitem{Agiwal_COMST_2018}
M.~Agiwal, A.~Roy, and N.~Saxena, ``{Next Generation 5G Wireless Networks: A Comprehensive Survey},'' \emph{IEEE Communications Surveys \& Tutorials}, vol.~18, no.~3, pp. 1617--1655, 2016.

\bibitem{Ullah_5Gusecases_IEEE_2019}
H.~Ullah, N.~Gopalakrishnan~Nair, A.~Moore, C.~Nugent, P.~Muschamp, and M.~Cuevas, ``{5G Communication: An Overview of Vehicle-to-Everything, Drones, and Healthcare Use-Cases},'' \emph{IEEE Access}, vol.~7, pp. 37\,251--37\,268, 2019.

\bibitem{Qadir_6Gusecases_2023}
Z.~Qadir, K.~N. Le, N.~Saeed, and H.~S. Munawar, ``{Towards 6G Internet of Things: Recent advances, use cases, and open challenges},'' \emph{ICT Express}, vol.~9, no.~3, pp. 296--312, 2023.

\bibitem{Shen_6Gresearch_ACM_2022}
L.-H. Shen, K.-T. Feng, and L.~Hanzo, ``{Five Facets of 6G: Research Challenges and Opportunities},'' \emph{ACM Comput. Surv.}, vol.~55, no.~11, pp. 1--39, 2023.

\bibitem{Rappaport_THz_IEEE_2019}
T.~S. Rappaport, Y.~Xing, O.~Kanhere, S.~Ju, A.~Madanayake, S.~Mandal, A.~Alkhateeb, and G.~C. Trichopoulos, ``{Wireless Communications and Applications Above 100 GHz: Opportunities and Challenges for 6G and Beyond},'' \emph{IEEE Access}, vol.~7, pp. {78\,729--78\,757}, 2019.

\bibitem{Ericsson_MobilityReport_2024}
\BIBentryALTinterwordspacing
P.~Jonsson~et al., ``{Ericsson Mobility Report 2024},'' {Ericsson}, Tech. Rep., 2024. [Online]. Available: \url{https://www.ericsson.com/en/reports-and-papers/mobility-report}
\BIBentrySTDinterwordspacing

\bibitem{ITU_2017}
{ITU}, ``{Minimum requirements related to technical performance for IMT-2020 radio interface(s)},'' {International Telecommunication Union}, ITU-R Reports M.2410-0, 2017.

\bibitem{Larsen_COMST_2019}
L.~M.~P. Larsen, A.~Checko, and H.~L. Christiansen, ``{A Survey of the Functional Splits Proposed for 5G Mobile Crosshaul Networks},'' \emph{IEEE Communications Surveys \& Tutorials}, vol.~21, no.~1, pp. 146--172, 2019.

\bibitem{Ceva_PentaGRAN}
\BIBentryALTinterwordspacing
``{Ceva PentaG, Open RAN Platform for Base Station and Radio}.'' [Online]. Available: \url{https://www.ceva-ip.com/product/ceva-pentag-ran/}
\BIBentrySTDinterwordspacing

\bibitem{Mitola_SDR_IEEE_1995}
J.~Mitola, ``{The software radio architecture},'' \emph{IEEE Communications Magazine}, vol.~33, no.~5, pp. 26--38, 1995.

\bibitem{Machado_SDR_IEEE_2015}
R.~G. Machado and A.~M. Wyglinski, ``{Software-Defined Radio: Bridging the Analog–Digital Divide},'' \emph{Proceedings of the IEEE}, vol. 103, no.~3, pp. 409--423, 2015.

\bibitem{Marvell_SDK}
\BIBentryALTinterwordspacing
``{Software (SDK) for OCTEON Multi-Core MIPS64 Processors}.'' [Online]. Available: \url{https://www.marvell.com/products/infrastructure-processors/software-development-kit.html}
\BIBentrySTDinterwordspacing

\bibitem{Quallcom_SDK}
\BIBentryALTinterwordspacing
``{Hexagon DSP SDK}.'' [Online]. Available: \url{https://developer.qualcomm.com/software/hexagon-dsp-sdk/dsp-processor}
\BIBentrySTDinterwordspacing

\bibitem{Huawei_ROI}
\BIBentryALTinterwordspacing
``{Huawei's RuralStar2.0 Shortens ROI Periods to Less than 5 Years},'' 2017. [Online]. Available: \url{https://www.huawei.com/en/news/2017/11/huawei-ruralstar2}
\BIBentrySTDinterwordspacing

\bibitem{Wesemann_IEEE_2023}
S.~Wesemann, J.~Du, and H.~Viswanathan, ``Energy efficient extreme mimo: Design goals and directions,'' \emph{IEEE Communications Magazine}, vol.~61, no.~10, pp. 132--138, 2023.

\bibitem{Blume_IEEE_2010}
O.~Blume, D.~Zeller, and U.~Barth, ``Approaches to energy efficient wireless access networks,'' in \emph{2010 4th International Symposium on Communications, Control and Signal Processing (ISCCSP)}, 2010, pp. 1--5.

\bibitem{Huawei_Power}
\BIBentryALTinterwordspacing
``{5G Power: Creating a green grid that slashes costs, emissions \& energy use},'' 2020. [Online]. Available: \url{https://www.huawei.com/en/huaweitech/publication/89/5g-power-green-grid-slashes-costs-emissions-energy-use}
\BIBentrySTDinterwordspacing

\bibitem{Zhang_TeraPool_2024}
Y.~Zhang, S.~Riedel, M.~Bertuletti, A.~Vanelli-Coralli, and L.~Benini, ``{TeraPool-SDR: An 1.89TOPS 1024 RV-Cores 4MiB Shared-L1 Cluster for Next-Generation Open-Source Software-Defined Radios},'' in \emph{{GLSVLSI '24: Proceedings of the Great Lakes Symposium on VLSI 2024}}, 2024, p. 86–91.

\bibitem{Peng_MMSE_ASIC_TCAS_2018}
G.~Peng, L.~Liu, S.~Zhou, S.~Yin, and S.~Wei, ``{A 1.58 Gbps/W 0.40 Gbps/mm2 ASIC Implementation of MMSE Detection for $128\times 8~64$ -QAM Massive MIMO in 65 nm CMOS},'' \emph{IEEE Transactions on Circuits and Systems I: Regular Papers}, vol.~65, no.~5, pp. 1717--1730, 2018.

\bibitem{Tang_MMSE_ASIC_JSSCC_2021}
W.~Tang, C.-H. Chen, and Z.~Zhang, ``{A 0.58-mm2 2.76-Gb/s 79.8-pJ/b 256-QAM Message-Passing Detector for a 128 × 32 Massive MIMO Uplink System},'' \emph{IEEE Journal of Solid-State Circuits}, vol.~56, no.~6, pp. 1722--1731, 2021.

\bibitem{Shahabuddin_MIMO_TVLSI_2021}
S.~Shahabuddin, I.~Hautala, M.~Juntti, and C.~Studer, ``{ADMM-Based Infinity-Norm Detection for Massive MIMO: Algorithm and VLSI Architecture},'' \emph{IEEE Transactions on Very Large Scale Integration (VLSI) Systems}, vol.~29, no.~4, pp. 747--759, 2021.

\bibitem{Guo_FFT_ASIC_TVLSI_2023}
Y.~Guo, Z.~Wang, Q.~Hong, H.~Luo, X.~Qiu, and L.~Liang, ``{A 60-Mode High-Throughput Parallel-Processing FFT Processor for 5G/4G Applications},'' \emph{IEEE Transactions on Very Large Scale Integration (VLSI) Systems}, vol.~31, no.~2, pp. 219--232, 2023.

\bibitem{Yang_FFT_ASIC_TCAS_2023}
C.~Yang, J.~Wu, S.~Xiang, L.~Liang, and L.~Geng, ``{A High-Throughput and Flexible Architecture Based on a Reconfigurable Mixed-Radix FFT With Twiddle Factor Compression and Conflict-Free Access},'' \emph{IEEE Transactions on Very Large Scale Integration (VLSI) Systems}, vol.~31, no.~10, pp. 1472--1485, 2023.

\bibitem{Zhang_MMSE_ASIC_ISSCC_2024}
Y.~Zhang, W.~Zhou, Y.~Zhang, H.~Ji, Y.~Huang, X.~You, and C.~Zhang, ``{BayesBB: A 9.6Gbps 1.61ms Configurable All-MessagePassing Baseband-Accelerator for B5G/6G Cell-Free Massive-MIMO in 40nm CMOS},'' in \emph{2024 IEEE International Solid-State Circuits Conference (ISSCC)}, vol.~67, 2024, pp. 48--50.

\bibitem{EdgeQ}
\BIBentryALTinterwordspacing
``{5G Meets AI, World's First Base Station on a Chip}.'' [Online]. Available: \url{https://www.edgeq.io/technology/}
\BIBentrySTDinterwordspacing

\bibitem{Picocom_PC802}
\BIBentryALTinterwordspacing
``{PC802 Unleashed}.'' [Online]. Available: \url{https://picocom.com/products/socs/pc802/}
\BIBentrySTDinterwordspacing

\bibitem{Marvell_Octeon10}
\BIBentryALTinterwordspacing
``{Data Processing Units, Empowering 5G carrier, enterprise and AI cloud data infrastructure}.'' [Online]. Available: \url{https://www.marvell.com/products/data-processing-units.html}
\BIBentrySTDinterwordspacing

\bibitem{Quallcom_X100}
\BIBentryALTinterwordspacing
``{How we Won the Acceleration Architecture Debate}.'' [Online]. Available: \url{https://www.qualcomm.com/news/onq/2023/03/how-we-won-the-acceleration-architecture-debate}
\BIBentrySTDinterwordspacing

\bibitem{Bertuletti_PUSCH_2023}
M.~Bertuletti, Y.~Zhang, A.~Vanelli-Coralli, and L.~Benini, ``{Efficient Parallelization of 5G-PUSCH on a Scalable RISC-V Many-Core Processor},'' in \emph{2023 Design, Automation \& Test in Europe Conference \& Exhibition (DATE)}, 2023, pp. 1--6.

\bibitem{RISCV_zfinx}
\BIBentryALTinterwordspacing
``{Announcing public review for RISC-V standard extensions Zfinx, Zdinx, Zhinx, and Zhinxmin}.'' [Online]. Available: \url{https://riscv.org/blog/2021/08/announcing-public-review-for-risc-v-standard-extensions-zfinx-zdinx-zhinx-and-zhinxmin/}
\BIBentrySTDinterwordspacing

\bibitem{Zaruba_snitch_2021}
F.~Zaruba, F.~Schuiki, T.~Hoefler, and L.~Benini, ``{Snitch: A Tiny Pseudo Dual-Issue Processor for Area and Energy Efficient Execution of Floating-Point Intensive Workloads},'' \emph{IEEE Transactions on Computers}, vol.~70, no.~11, pp. 1845--1860, 2021.

\bibitem{Ingemarsson_inversion_2015}
C.~Ingemarsson and O.~Gustafsson, ``{On fixed-point implementation of symmetric matrix inversion},'' in \emph{2015 European Conference on Circuit Theory and Design (ECCTD)}, 2015, pp. 1--4.

\bibitem{Mach_fpnew_2021}
S.~Mach, F.~Schuiki, F.~Zaruba, and L.~Benini, ``{FPnew: An Open-Source Multiformat Floating-Point Unit Architecture for Energy-Proportional Transprecision Computing},'' \emph{IEEE Transactions on Very Large Scale Integration (VLSI) Systems}, vol.~29, no.~4, pp. 774--787, 2021.

\bibitem{Bertaccini_Minifloats_2024}
L.~Bertaccini, G.~Paulin, M.~Cavalcante, T.~Fischer, S.~Mach, and L.~Benini, ``{MiniFloats on RISC-V Cores: ISA Extensions with Mixed-Precision Short Dot Products},'' \emph{IEEE Transactions on Emerging Topics in Computing}, pp. 1--16, 2024.

\bibitem{Shahabuddin_MIMO_NorCAS_2020}
S.~Shahabuddin, M.~H. Islam, M.~S. Shahabuddin, M.~A. Albreem, and M.~Juntti, ``{Matrix Decomposition for Massive MIMO Detection},'' in \emph{2020 {IEEE} Nordic Circuits and Systems Conference ({NorCAS})}, 2020, pp. 1--6.

\bibitem{Riedel_MemPool_2023}
S.~Riedel, M.~Cavalcante, R.~Andri, and L.~Benini, ``{MemPool: A Scalable Manycore Architecture With a Low-Latency Shared L1 Memory},'' \emph{IEEE Transactions on Computers}, vol.~72, no.~12, pp. 3561--3575, 2023.

\bibitem{Bertuletti_Barriers_2023}
M.~Bertuletti, S.~Riedel, Y.~Zhang, A.~Vanelli-Coralli, and L.~Benini, ``{Fast Shared-Memory Barrier Synchronization a 1024-Cores RISC-V Many-Core Cluster},'' in \emph{Embedded Computer Systems: Architectures, Modeling, and Simulation: 23rd International Conference, SAMOS 2023, Samos, Greece, July 2–6, 2023, Proceedings}, 2023, p. 241–254.

\bibitem{3GPP_TS_R17_description}
{3GPP}, ``{NR and NG-RAN Overall Description},'' {3rd Generation Partnership Project (3GPP)}, Technical Specification (TS) 38.211, 2017, {Release 17}.

\bibitem{amor_toc_2022}
H.~B. Amor, C.~Bernier, and Z.~Přikryl, ``A risc-v isa extension for ultra-low power iot wireless signal processing,'' \emph{IEEE Transactions on Computers}, vol.~71, no.~4, pp. 766--778, 2022.

\bibitem{Chen_MIMO_TVLSI_2015}
X.~Chen, A.~Minwegen, S.~B. Hussain, A.~Chattopadhyay, G.~Ascheid, and R.~Leupers, ``{Flexible, Efficient Multimode MIMO Detection by Using Reconfigurable ASIP},'' \emph{IEEE Transactions on Very Large Scale Integration (VLSI) Systems}, vol.~23, no.~10, pp. 2173--2186, 2015.

\bibitem{Kultala_Lord_TCAS_2019}
H.~Kultala, T.~Viitanen, H.~Berg, P.~Jääskeläinen, J.~Multanen, M.~Kokkonen, K.~Raiskila, T.~Zetterman, and J.~Takala, ``{LordCore: Energy-Efficient OpenCL-Programmable Software-Defined Radio Coprocessor},'' \emph{IEEE Transactions on Very Large Scale Integration (VLSI) Systems}, vol.~27, no.~5, pp. 1029--1042, 2019.

\bibitem{Fu_DREAM_TCAS_2022}
Y.~Fu, K.~Chen, W.~Song, G.~He, S.~Shen, H.~Wang, C.~Zhang, and L.~Li, ``{A DSP-Purposed REconfigurable Acceleration Machine (DREAM) for High Energy Efficiency MIMO Signal Processing},'' \emph{IEEE Transactions on Circuits and Systems I: Regular Papers}, vol.~70, no.~2, pp. 952--965, 2023.

\bibitem{Attari_VectorASIP_TCAS_2022}
M.~Attari, L.~Ferreira, L.~Liu, and S.~Malkowsky, ``{An Application Specific Vector Processor for Efficient Massive MIMO Processing},'' \emph{IEEE Transactions on Circuits and Systems I: Regular Papers}, vol.~69, no.~9, pp. 3804--3815, 2022.

\bibitem{Castaneda_MMSE_ASIC_ESSCIRC_2022}
O.~Castañeda, L.~Benini, and C.~Studer, ``{A 283 pJ/b 240 Mb/s Floating-Point Baseband Accelerator for Massive MU-MIMO in 22FDX},'' in \emph{ESSCIRC 2022- IEEE 48th European Solid State Circuits Conference (ESSCIRC)}, 2022, pp. 357--360.

\bibitem{Chen_DXT501_CoolChips_2022}
Y.~Chen, L.~Liu, X.~Feng, and J.~Shi, ``{DXT501: An SDR-Based Baseband MP-SoC for Multi-Protocol Industrial Wireless Communication},'' in \emph{2022 IEEE Symposium in Low-Power and High-Speed Chips (COOL CHIPS)}, 2022, pp. 1--6.

\bibitem{ax800}
\BIBentryALTinterwordspacing
``{NVIDIA AX800 Delivers High-Performance 5G vRAN and AI Services on One Common Cloud Infrastructure}.'' [Online]. Available: \url{https://developer.nvidia.com/blog/nvidia-ax800-delivers-high-performance-5g-vran-and-ai-services-on-one-common-cloud-infrastructure/}
\BIBentrySTDinterwordspacing

\bibitem{Xilinx_RFSoC}
\BIBentryALTinterwordspacing
``{Zynq UltraScale+ RFSoC The Industry’s Only Single-Chip Adaptable Radio Platform}.'' [Online]. Available: \url{https://www.xilinx.com/products/silicon-devices/soc/rfsoc.html}
\BIBentrySTDinterwordspacing

\bibitem{tarver_gpu_2021}
C.~Tarver, M.~Tonnemacher, H.~Chen, J.~Zhang, and J.~R. Cavallaro, ``{GPU}-{Based}, {LDPC} {Decoding} for {5G} and {Beyond},'' \emph{IEEE Open Journal of Circuits and Systems}, vol.~2, pp. 278--290, 2021.

\bibitem{kundu_gpu_2023}
L.~Kundu, X.~Lin, E.~Agostini, V.~Ditya, and T.~Martin, ``Hardware acceleration for open radio access networks: A contemporary overview,'' \emph{IEEE Communications Magazine}, pp. 1--7, 2023.

\bibitem{kelkar_aerial_2021}
A.~Kelkar and C.~Dick, ``{NVIDIA} {Aerial} {GPU} {Hosted} {AI}-on-{5G},'' in \emph{2021 {IEEE} 4th {5G} {World} {Forum} ({5GWF})}, 2021, pp. 64--69.

\bibitem{hoydis_sionna_2022}
\BIBentryALTinterwordspacing
J.~Hoydis, S.~Cammerer, F.~A. Aoudia, A.~Vem, N.~Binder, G.~Marcus, and A.~Keller, ``Sionna: {An} {Open}-{Source} {Library} for {Next}-{Generation} {Physical} {Layer} {Research},'' 2022. [Online]. Available: \url{https://arxiv.org/abs/2203.11854}
\BIBentrySTDinterwordspacing

\bibitem{wu_gpu_2014}
M.~Wu, B.~Yin, G.~Wang, C.~Studer, and J.~R. Cavallaro, ``{GPU} {Acceleration} of a {Configurable} {N}-{Way} {MIMO} {Detector} for {Wireless} {Systems},'' \emph{Journal of Signal Processing Systems}, vol.~76, no.~2, pp. 95--108, 2014.

\bibitem{li_gpu_2015}
K.~Li, B.~Yin, M.~Wu, J.~R. Cavallaro, and C.~Studer, ``{Accelerating massive MIMO uplink detection on GPU for SDR systems},'' in \emph{2015 IEEE Dallas Circuits and Systems Conference (DCAS)}, 2015, pp. 1--4.

\bibitem{ARM_NeoverseN2}
A.~Pellegrini, ``{Arm Neoverse N2: Arm’s 2nd generation high performance infrastructure CPUs and system IPs},'' in \emph{2021 IEEE Hot Chips 33 Symposium (HCS)}, 2021, pp. 1--27.

\end{thebibliography}

\begin{IEEEbiography}[{\includegraphics[width=1in,height=1.25in,keepaspectratio,clip]{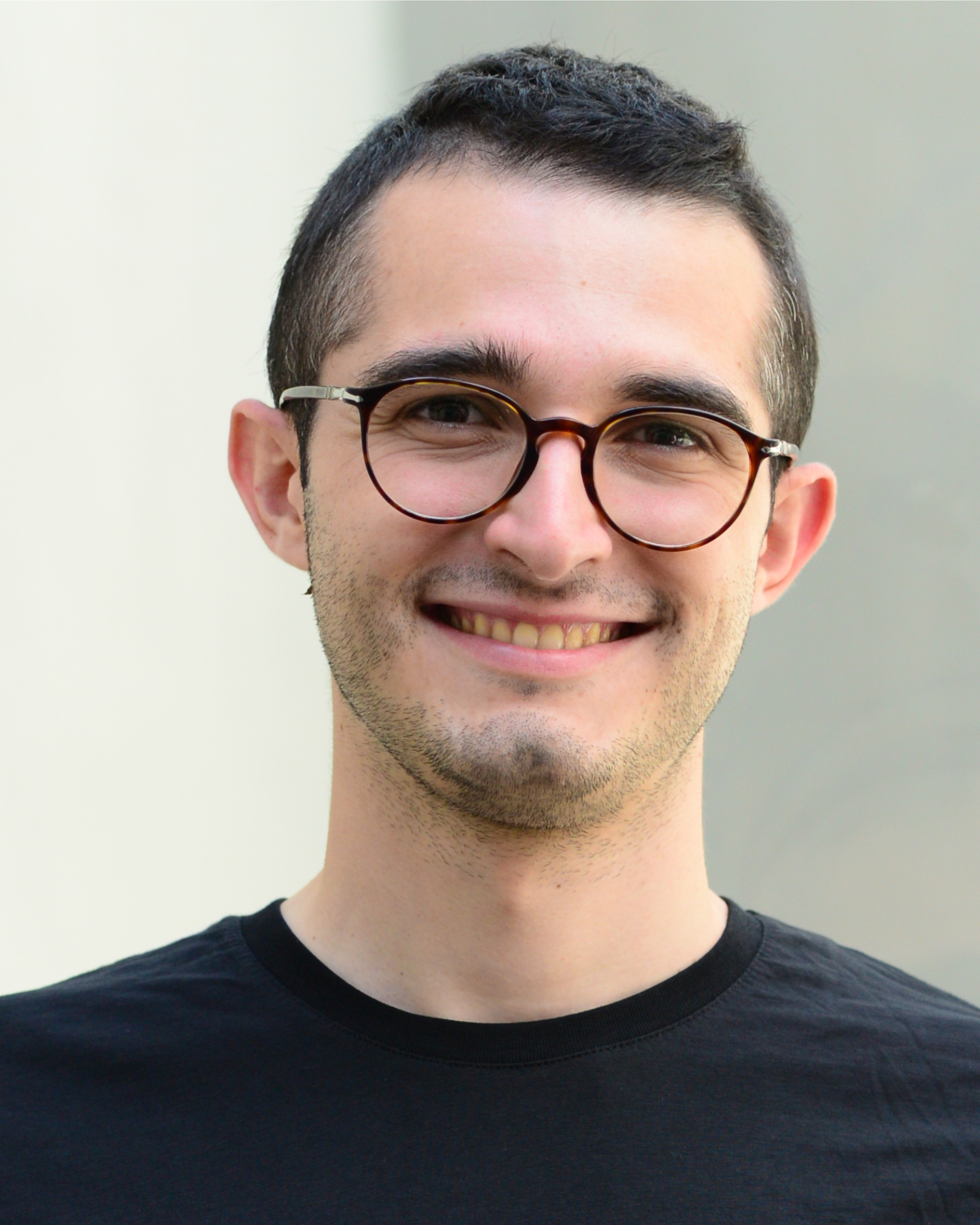}}]{Marco Bertuletti}
received the B.Sc. and M.Sc. degree in Electrical Engineering in Politecnico di Milano, Milano, Italy. He is currently pursuing his Ph.D. at ETH, Zurich, Switzerland, in the Integrated Systems Laboratory (IIS). His main interests are in the design of multi and many-core clusters of RISC-V processors for next-generation telecommunications.
\end{IEEEbiography}

\begin{IEEEbiography}[{\includegraphics[width=1in,height=1.25in,keepaspectratio,clip]{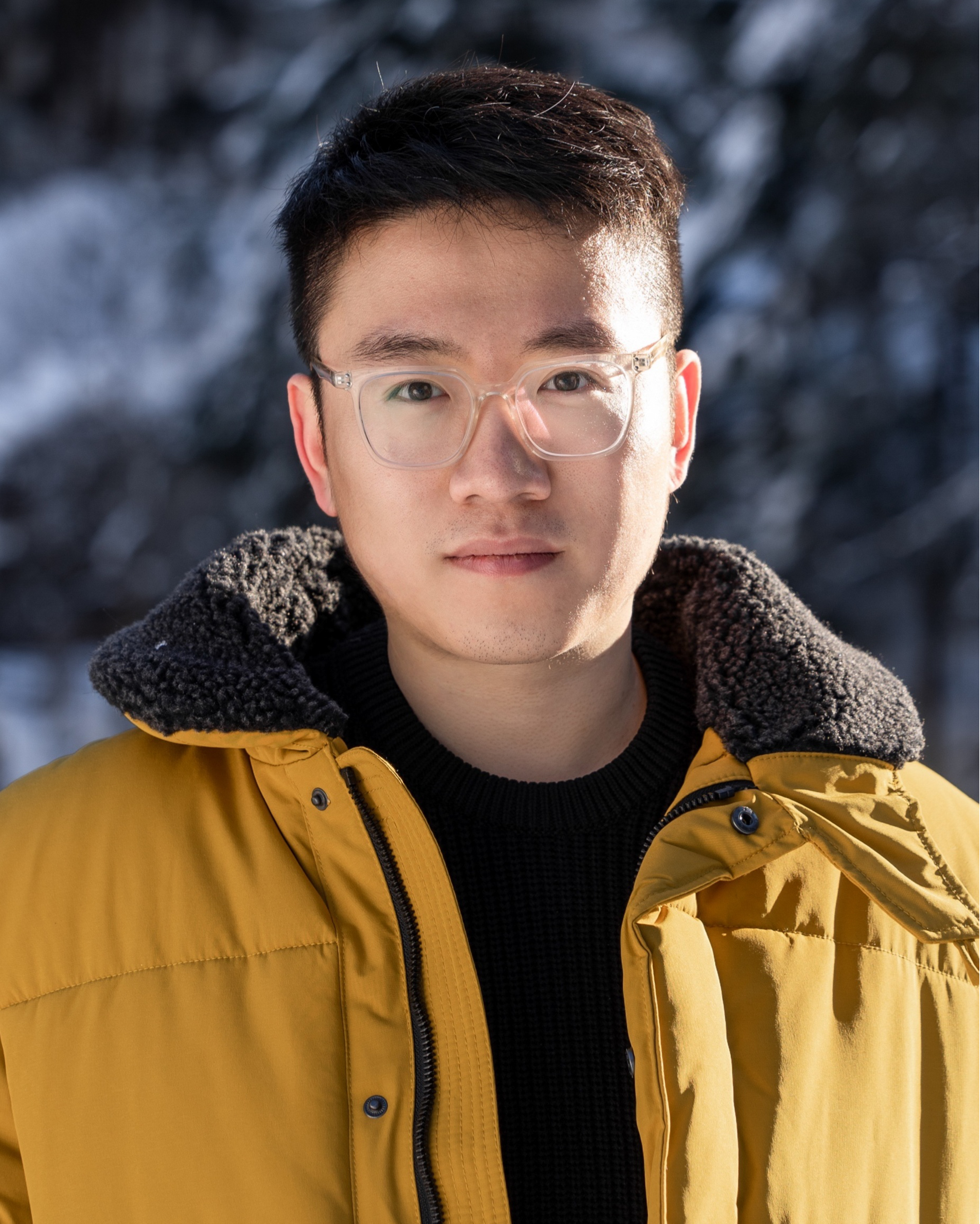}}]{Yichao Zhang}
received the M.Sc. degree in Electronic Science from Nanyang Technological University, Singapore, and the B.Eng. degree in Integrated Circuit Design from Heilongjiang University, China, in 2017 and 2015. He served in the physical VLSI design at Cadence Design Systems and MediaTek Singapore until 2021. He is currently pursuing a Ph.D. at ETH Zurich, Switzerland, in the Integrated Systems Laboratory (IIS). His research focuses on physically feasible, many-core RISC-V architectures, parallel computing, and SIMD processing.
\end{IEEEbiography}

\begin{IEEEbiography}
[{\includegraphics[width=1in,height=1.25in,keepaspectratio,clip]{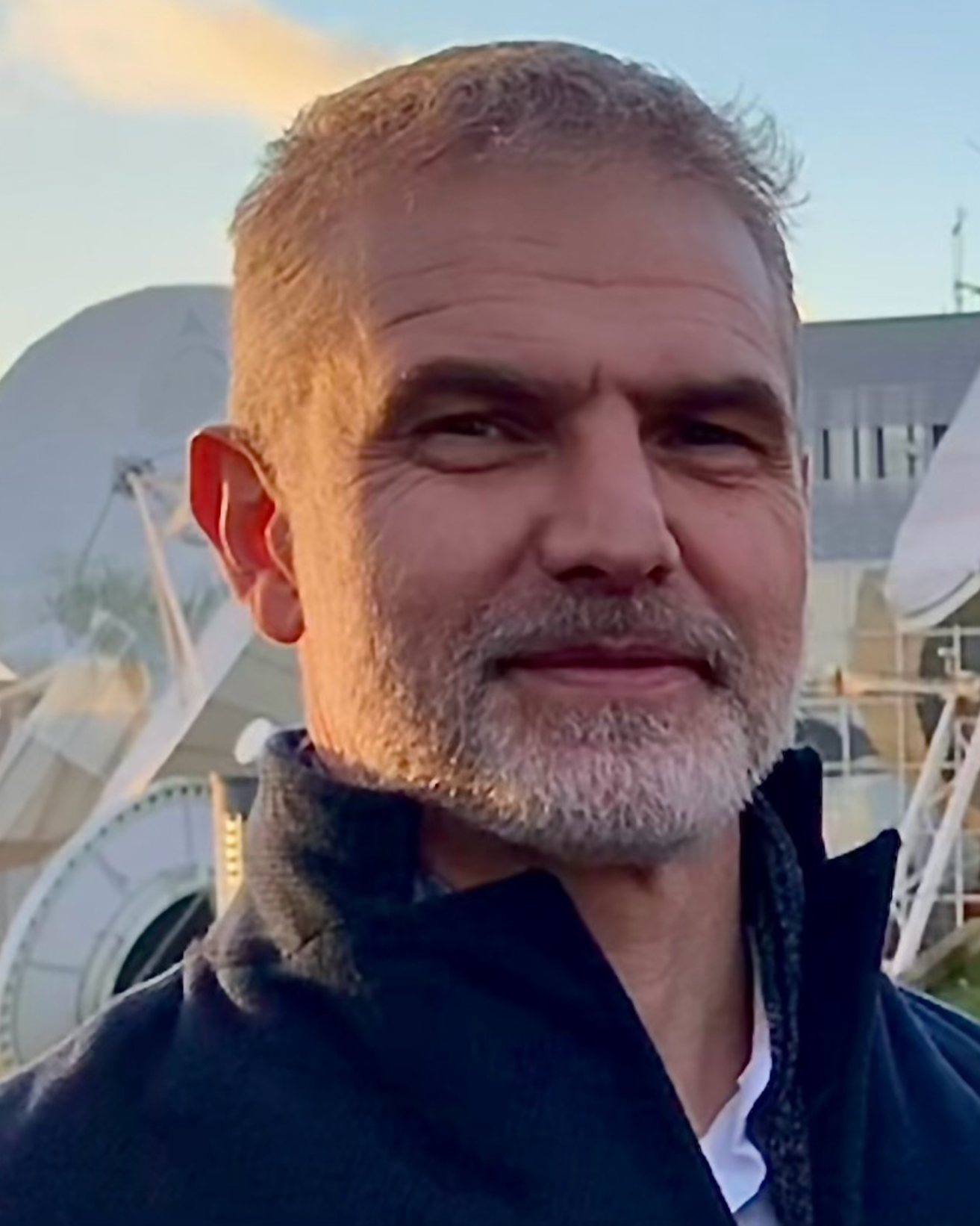}}]{Alessandro Vanelli-Coralli}
received the Dr.Ing. degree in electronics engineering and the Ph.D. degree in electronics and computer science from the University of Bologna, Italy, in 1991 and 1996, respectively, where he is currently a full Professor. Since 2022 he is also a Senior Scientist - ETH Zurich. From 2003 to 2005, he was a Visiting Scientist with Qualcomm Inc., San Diego, CA, USA. He participates in national and international research projects on wireless and satellite communication systems and he has been a Project Coordinator and scientific responsible for several European Space Agency and European Commission funded projects. He is currently the Responsible for the Vision and Research Strategy task force of the Networld2020 SatCom Working Group. Dr. Vanelli-Coralli has served in the organisation committees of scientific conferences. He is co-recipient of several Best Paper Awards and he is the recipient of the 2019 IEEE Satellite Communications Technical Recognition Award.
\end{IEEEbiography}

\begin{IEEEbiography}[{\includegraphics[width=1in,height=1.25in,keepaspectratio,clip]{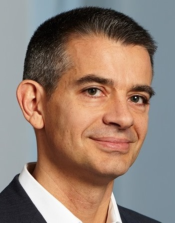}}]{Luca Benini}
holds the chair of digital Circuits and systems at ETHZ and is Full Professor at the Università di Bologna. He received a PhD from Stanford University. His research interests are in energy-efficient parallel computing systems, smart sensing micro-systems and machine learning hardware. He is a fellow of the ACM, a member of the Academia Europaea and of the Italian Academy of Engineering and Technology. 
\end{IEEEbiography}

\end{document}